\documentclass[twocolumn]{aastex62}
\usepackage{float}
\usepackage{amssymb,amsmath,natbib,gensymb}
\usepackage{color}
\usepackage{ulem}
\usepackage{latexsym}
\usepackage{wasysym}

\shorttitle{The rapidly declining Type~I SN~2019bkc}
\shortauthors{Chen et al.}
\begin{document}

\title{The Most Rapidly Declining Type I Supernova 2019bkc/ATLAS19dqr}
\author[0000-0003-0853-6427]{Ping Chen}\affil{Kavli Institute for Astronomy and Astrophysics, Peking University, Yi He Yuan Road 5, Hai Dian District, Beijing 100871, China.}\affil{Department of Astronomy, School of Physics, Peking University, Yi He Yuan Road 5, Hai Dian District, Beijing 100871, China}
\author[0000-0002-1027-0990]{Subo Dong}\affil{Kavli Institute for Astronomy and Astrophysics, Peking University, Yi He Yuan Road 5, Hai Dian District, Beijing 100871, China.}
\author[0000-0002-5571-1833]{M. D. Stritzinger} \affil{Department of Physics and Astronomy, Aarhus University, Ny Munkegade 120, DK-8000 Aarhus C, Denmark} 
\author{Simon Holmbo}\affil{Department of Physics and Astronomy, Aarhus University, Ny Munkegade 120, DK-8000 Aarhus C, Denmark} 
\author{Jay Strader}
\affil{Center for Data Intensive and Time Domain Astronomy, Department of Physics and Astronomy, Michigan State University, East Lansing, MI 48824, USA}
\author{C. S. Kochanek} \affil{Department of Astronomy, The Ohio State University, 140 West 18th Avenue, Columbus, OH 43210, USA}\affil{Center for Cosmology and AstroParticle Physics, The Ohio State University, 191 W. Woodruff Ave., Columbus, OH 43210, USA}
\author{Eric W. Peng}\affil{Department of Astronomy, School of Physics, Peking University, Yi He Yuan Road 5, Hai Dian District, Beijing 100871, China}\affil{Kavli Institute for Astronomy and Astrophysics, Peking University, Yi He Yuan Road 5, Hai Dian District, Beijing 100871, China.}
\author{S. Benetti}\affil{INAF - Osservatorio Astronomico di Padova, Vicolo dell'Osservatorio 5, I-35122 Padova, Italy}
\author{D. Bersier}\affil{Astrophysics Research Institute, Liverpool John Moores University, 146 Brownlow Hill, Liverpool L3 5RF, UK}
\author{Sasha Brownsberger}\affil{Department of Physics, Harvard University, Cambridge, MA 02138 , USA}
\author{David A. H. Buckley}\affil{South African Astronomical Observatory, PO Box 9, Observatory 7935, Cape Town, South Africa}
\author{Mariusz Gromadzki}\affil{Warsaw University Astronomical Observatory, Al. Ujazdowskie 4, 00-478 Warszawa, Poland}
\author{Shane Moran}\affil{Tuorla Observatory, Department of Physics and Astronomy, FI-20014, University of Turku, Finland}\affil{Nordic Optical Telescope, Apartado 474, E-38700 Santa Cruz de La Palma, Spain}
\author{A. Pastorello}\affil{INAF - Osservatorio Astronomico di Padova, Vicolo dell'Osservatorio 5, I-35122 Padova, Italy}
\author[0000-0001-8525-3442]{Elias Aydi}\affil{Center for Data Intensive and Time Domain Astronomy, Department of Physics and Astronomy, Michigan State University, East Lansing, MI 48824, USA}
\author{Subhash Bose}\affil{Kavli Institute for Astronomy and Astrophysics, Peking University, Yi He Yuan Road 5, Hai Dian District, Beijing 100871, China.}
\author[0000-0002-7898-7664]{Thomas Connor}\affil{The Observatories of the Carnegie Institution for Science, 813 Santa Barbara St., Pasadena, CA 91101, USA}
\author{K. Boutsia}\affil{Las Campanas Observatory, Carnegie Observatories, Casilla 601, La Serena, Chile}
\author{F. Di Mille}\affil{Las Campanas Observatory, Carnegie Observatories, Casilla 601, La Serena, Chile}
\author{N. Elias-Rosa}\affil{Institute of Space Sciences (ICE-CSIC), Campus UAB, Carrer de Can Magrans S/N, 08193 Barcelona, Spain}\affil{Institut d'Estudis Espacials de Catalunya (IEEC), c/Gran Capit\'a 2-4, Edif. Nexus 201, 08034 Barcelona, Spain}
\author{K. Decker French}\altaffiliation{Hubble Fellow}\affil{The Observatories of the Carnegie Institution for Science, 813 Santa Barbara St., Pasadena, CA 91101, USA} \author[0000-0001-9206-3460]{Thomas~W.-S.~Holoien}\altaffiliation{Carnegie Fellow}\affiliation{The Observatories of the Carnegie Institution for Science, 813 Santa Barbara St., Pasadena, CA 91101, USA}
\author{Seppo Mattila}\affil{Tuorla Observatory, Department of Physics and Astronomy, FI-20014, University of Turku, Finland}
\author{B.~J. Shappee}\affil{Institute for Astronomy, University of Hawaii, 2680 Woodlawn Drive, Honolulu, HI 96822, USA}
\author[0000-0002-2718-9996]{Antony~A. Stark}\affil{Center for Astrophysics | Harvard \& Smithsonian, 60 Garden St.; Cambridge MA 02138 USA}
\author{Samuel J. Swihart}\affil{Center for Data Intensive and Time Domain Astronomy, Department of Physics and Astronomy, Michigan State University, East Lansing, MI 48824, USA}

\correspondingauthor{Subo Dong}
\email{dongsubo@pku.edu.cn}

\begin{abstract}

We report observations of the hydrogen-deficient supernova (SN) 2019bkc/ATLAS19dqr. With $B$- and $r$-band decline  between peak and 10 days post peak of $\Delta{m_{10}(B)}=5.24\pm0.07$ mag and  $\Delta{m_{10}(r)}=3.85\pm0.10$ mag, respectively,  SN~2019bkc is the most rapidly declining SN I discovered so far.  While its  closest matches are the rapidly declining SN~2005ek and SN~2010X, the LCs and spectra of SN~2019bkc show some unprecedented characteristics. SN~2019bkc appears ``hostless,'' with no identifiable host galaxy near its location, although it may be associated with the galaxy cluster MKW1 at $z=0.02$. We evaluate a number of existing models of fast-evolving SNe, and we find that none of them can satisfactorily explain all aspects of SN~2019bkc observations.

\end{abstract}

\keywords{supernovae: general $-$ supernovae: individual: (SN~2019bkc/ATLAS19dqr)}

\section{INTRODUCTION}

Modern wide-field time-domain surveys have significantly expanded the discovery space of the transient universe, and in particular, 
surveys with higher cadences are finding an increasing number of transients that rise and/or fall much faster than ordinary supernovae (SNe). A number of rapidly evolving transients with spectroscopic data nominally belong to the H-deficient SN I class. Some well-studied examples include SN~2002bj \citep{Poznanski2010},  SN~2010X \citep{Kasliwal2010}, SN~2005ek \citep{Drout2013}, OGLE-2013-SN-079  \citep{Inserra2015}, and iPTF 14gqr \citep{De2018}. 
These rapidly evolving objects likely have small ejecta masses (a few tenths of $M_\odot$), and broadly speaking, they can be divided into those with He lines (e.g., SN~2002bj) and without (e.g., SN~2010X), making them nominally Type~Ib or Type~Ic, respectively. In a few cases, fast-evolving transients show clear evidence of interaction with He-rich circumstellar material \citep[CSM; e.g.,][]{Pastorello2015}.  Most of the known fast-evolving  transients have been identified in large imaging surveys and lack spectroscopic follow-up \citep[e.g.,][]{Drout2014, Pursiainen2018, Rest2018}. The theoretical interpretation of these fast-evolving  SNe~I is inconclusive, and include scenarios  ranging from  He-shell detonations on the surface of sub-Chandrasekhar mass white dwarf (WD) stars (the so-called SNe~``.Ia''; \citealt{Bildsten2007, Shen2010}) or possibly WDs that are disrupted in an accretion-induced collapse (AIC; see, e.g., \citealt{Dessart2006,Darbha2010}), an ultra-stripped massive star explosion \citep[see, e.g.,][]{Tauris2013a}, core-collapse supernovae (ccSNe) with fallback \citep[see, e.g.,][]{Moriya2010}, fallback accretion of while dwarf--neutron star (WD--NS) mergers \citep{Margalit2016}, He-stars with an extended envelope \citep{Kleiser2014}, or being powered by highly magnetized NSs \citep{Yu2015}.

\begin{figure*}
\centerline{\includegraphics[width=18cm,height=6cm]{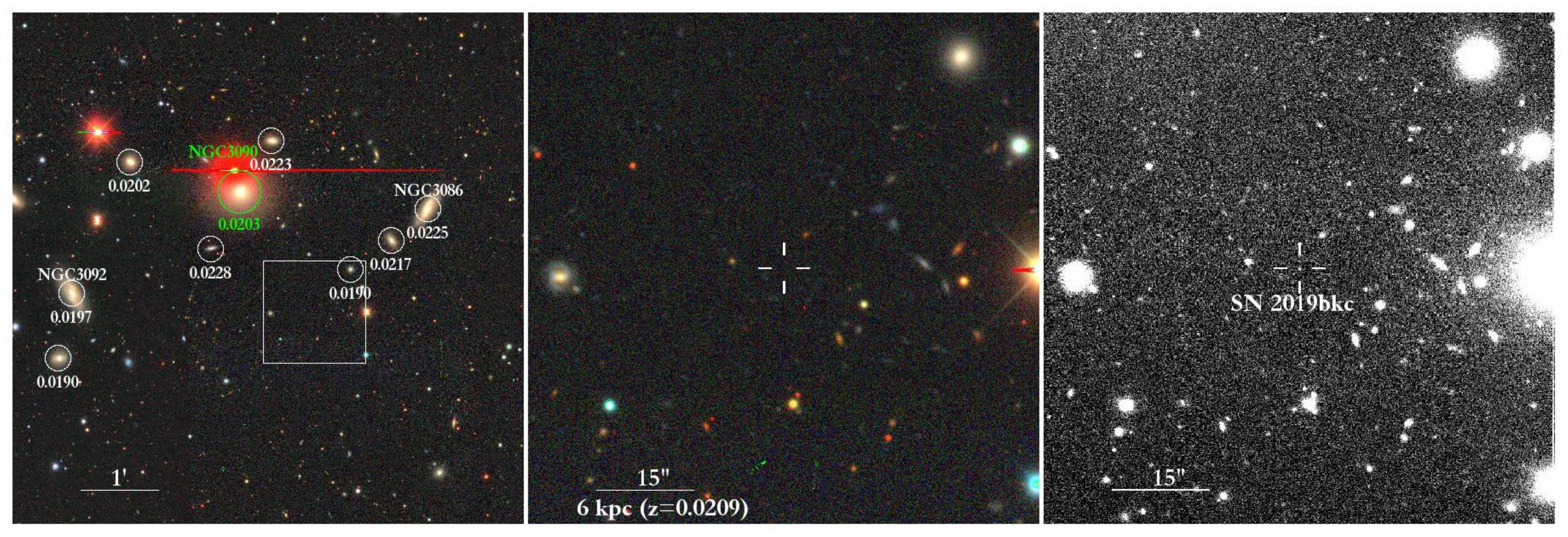}}
\caption{Left panel: composite Dark Energy Camera Legacy Survey (DECaLS) $grz$ image of the field of SN~2019bkc. The redshifts of the cD galaxy NGC 3090 along with NGC 3092, NGC 3086 and other members of the galaxy cluster MKW1 are labeled.  Middle panel: zoom in on the boxed region from the Left panel centered on SN~2019bkc whose position is indicated by the white cross. Right panel: a deep $i$-band Nordic Optical Telescope (NOT) image including SN~2019bkc of the same region.}
\label{fig:host}
\end{figure*}

Here we report observations of the rapidly evolving  Type~I SN~2019bkc. SN~2019bkc exploded in an isolated environment and exhibited the most rapid post-peak photometric decline yet observed.

\begin{figure*}
\includegraphics[width=18cm]{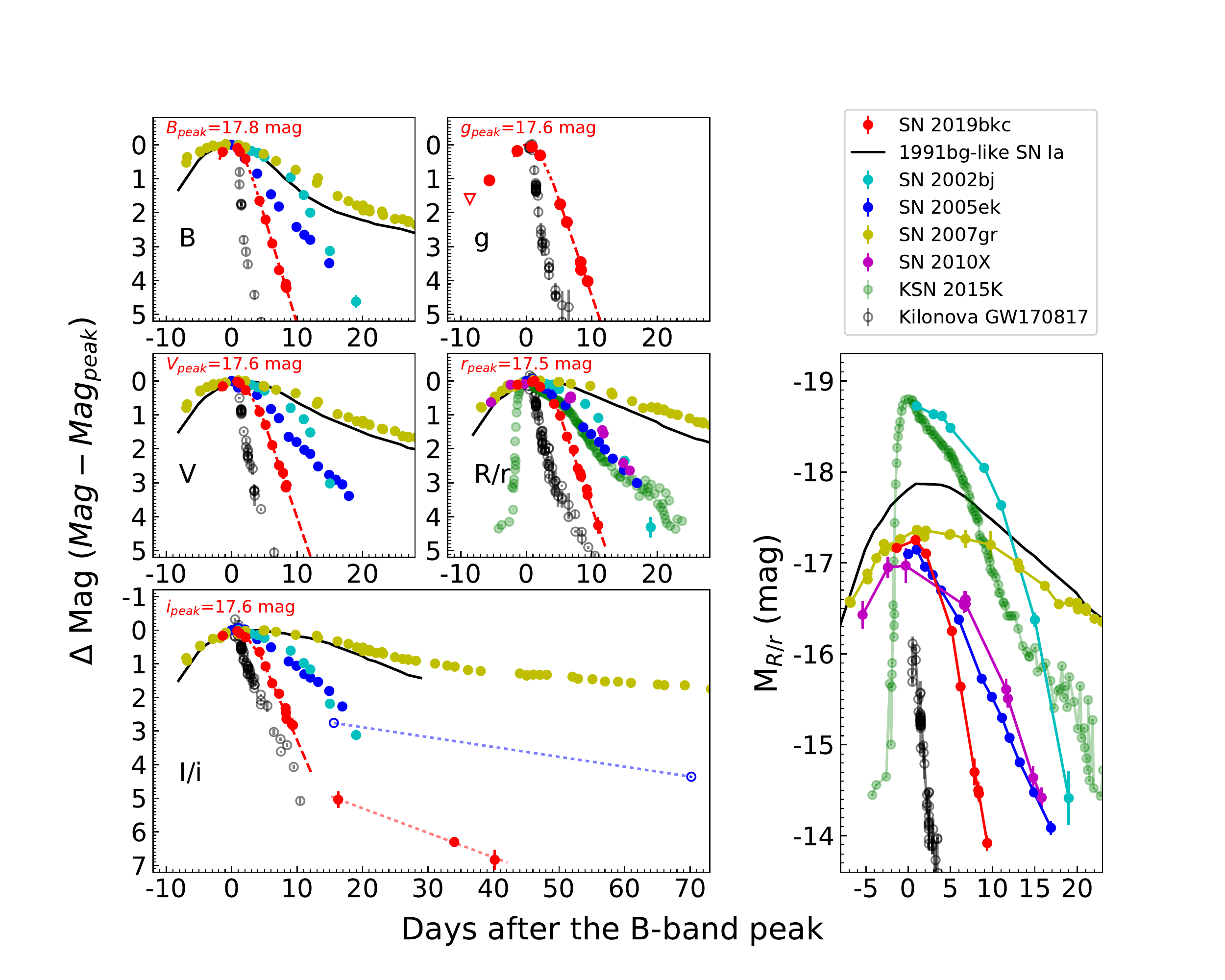}
\caption{$B$-, $V$-, $g$-, $R/r$-, and $I/i$-band LCs of SN~2019bkc compared with the  fast-evolving Type~I~SNe 2002bj, 2005ek, and 2010X, the fastest-declining 1991bg-like Type~Ia, the Type~Ic~SN~2007gr, and the fast transient KSN 2015K (in Kepmag) found by {\it Kepler} and the kilonova GW 170817. The left panels normalize the LCs to the peak brightness, while the right panel shows the absolute $R/r$-band magnitudes. The dashed lines represent the third-order polynomial fit to the LCs used to estimate the peak (displayed up to $+4$ days) and the linear fits between $+4$ and $+10$ days used to estimate the decline-rate parameters in $\S$~\ref{photometric_results}. The red and blue dotted line is the best-fit straight line for the late-time $i$-band LC of 2019bkc and 2005ek, respectively.}
\label{fig:lcs_compare}
\end{figure*}

 \section{Discovery}
 
ATLAS19dqr/SN~2019bkc (R.A.$=10^h00^m22\fs544$, decl.$=-03\degr01\arcmin12\farcs64$) was discovered by the Asteroid Terrestrial-impact Last Alert System (ATLAS; \citealt{Tonry2018}) on 2019 March 2.43 UT 
 in the ALTAS ``orange'' filter at an apparent magnitude of $m_o=17.96\pm0.06$\,mag \citep{Tonry2019}. Following the practice of the ATLAS project we refer to the source as SN~2019bkc.
 
 As can be seen in Figure~\ref{fig:host}, SN~2019bkc is ``hostless,'' as no source is detected at its location down to $g\approx26.2$ mag ($3\,\sigma$ upper limit) in deep archived images available from NOAO Data Lab\footnote{\url{https://datalab.noao.edu/}} obtained with DECam on CTIO-4m telescope under program 2017A-0060. 

A hostless transient can be an object in the Galaxy (e.g., a cataclysmic variable). We apply the classic Baade--Wesselink method \citep{Wesselink1946} to make an order-of-magnitude assessment based on basic observed properties (derived from light curves (LCs) and spectra discussed below) on whether it is in the Galaxy. We can estimate its physical size at peak by using the velocity (on the order of $\sim$10000 km\,s$^{-1}$ from the width of absorption line in the spectra) and rise timescale ($\sim$10 days) from the LC, which is $\sim$10000 km\,s$^{-1}$ $\times$ 10 days $\sim 10^{15}$ cm. With peak temperature on the order of $\sim$10000\,K and apparent magnitude ($\sim$18 mag), its angular radius is estimated to be $\sim 5\times10^{-7}$ arcsec at peak.  Therefore, its distance is on the order of $\sim$100\,Mpc, suggesting that SN 2019bkc is most likely extragalactic.

Its apparent location is within the poor cluster MKW1 \citep{Morgan1975}, and it is located 217\arcsec\,from the cD galaxy NGC~3090 of the cluster. In Appendix~\ref{sec:environment}, we estimate the chance of alignment of SN~2019bkc with galaxies in the field of view and conclude that it is most likely associated with MKW1. The averaged radial velocity of MKW1 members  within an $0.5h^{-1}$ Mpc radius of NGC~3090 is $6252\pm95$ km\,s$^{-1}$ \citep{Koranyi2002}, and we adopt $z=0.0209\pm0.0003$ as the redshift of SN~2019bkc. We stress that there is no direct measurement definitively establishing the redshift of SN~2019bkc, and we note the associated caveats in the text. Correcting for the infall velocity of the Local Group toward the Virgo Cluster yields a luminosity distance of $d_L=86.5\pm4.3$\,Mpc assuming a $\Lambda$CDM cosmology ($H_0 = 73\,{\rm km\,s}^{-1}\,{\rm Mpc}^{-1}$, $\Omega_m=0.27$ and $\Omega_{\Lambda}=0.73$) and $\sigma_{cz}=300~{\rm km\,s^{-1}}$ to account for peculiar velocities.

\section{Follow-up Observations}
\label{observations}

We followed up 2019bkc as a candidate of a complete, magnitude-limited sample of ccSNe. We took the first epoch of imaging on 2019 March 3.53 UT using the Sinistro cameras mounted on a 1 m telescope of Las Cumbres Observatory Global Telescope Network (LCOGT; \citealt{Brown2013}) and our first spectrum using the Goodman High Throughput Spectrograph (GTHS; \citealt{Clemens2004}) on 4.1 m Southern Astrophysical Research (SOAR) telescope on 2019 March 5.27 UT. The peculiar SOAR spectrum and the rapid post-peak decline of the SN motivated us to carry out an extensive follow-up campaign of the SN. On 2019 March 6.44 UT, \citet{Prentice2019} reported a spectrum taken on 2019 March 4.92 UT and remarked that SN~2019bkc was an ``unknown transient at an unknown redshift.'' The extended Public ESO Spectroscopic Survey for Transient Objects (ePESSTO; \citealt{PESSTO}) classified it as a likely SN~Ic based a spectrum taken on 2019 March 6 UT \citep{2019bkcAtel}.
 
We obtained optical/near-infrared (NIR) imaging with the 1 m LCOGT,  A Novel Dual Imaging CAMera (ANDICAM; \citealt{DePoy2003}) on the 1.3 m Small \& Moderate Aperture Research Telescope System (SMARTS; \citealt{Subasavage2010}), the Wide Field CAMera (WFCAM; \citealt{Casali2007}) mounted on the 3.8 m United Kingdom Infrared Telescope (UKIRT), the 2.56 m Nordic Optical Telescope (NOT), 2 m Liverpool Telescope (LT), and the 6.5 m \textit{Magellan} telescopes. We detected SN from $-1.4$ days to $+$40.2 days relative to the epoch of $B$-band peak (see below). We obtained four visual-wavelength spectra on +0.4 day (SOAR), +7.1 days (The Southern African Large Telescope, SALT), +7.1 days (NOT), and +10.1 days (NOT). Descriptions of the data acquisition and reduction are provided in Appendices~\ref{sec:phot_obs} and \ref{sec:spec_obs}.  

\subsection{LC Evolution}
\label{photometric_results}

The optical LCs are well described by third-order polynomials except for the late-time $i-$band data after $+16$ days. We measure the time and magnitude of the peak using the best-fit third-order polynomials (see Table~\ref{lc_paremeters}), and the parameter uncertainties are estimated with the Markov chain Monte Carlo (MCMC) method. MCMC method is also used to estimate uncertainties for other parameters in Table~\ref{lc_paremeters} except for the absolute magnitudes for which the uncertainties are dominated by uncertainties of distance to SN 2019bkc. The $B$-band peak time of JD $2458547.39 \pm 0.04$ (2019 Mar 04.90 UT) is used as the reference epoch for the SN phase. 
 
There was an ALTAS upper limit on  $-6.4$ days at $m_o>19.33$\,mag preceding the $-2.5$ days detection  \citep{Tonry2019}. We also access data taken by the Zwicky Transient Facility (ZTF; \citealt{ZTF}) through Lasair \citep{Lasair}. There was a detection on $-5.6$ days at $m_g = 18.69\pm0.11$\,mag and a previous upper limit on $-8.6$ days of $m_g > 19.24$\,mag . 

The $g$-band LC of SN~2019bkc (Figure~\ref{fig:lcs_compare}) rose by $1$\,mag from detection to peak in $\sim$6 days. In contrast, the $g$-band flux dropped by $\sim 2$\,mag by $+$6 days. In each optical band, between $\sim$4 and 10 days, the LCs show rapid exponential flux declines $f_{\lambda} \propto \exp(-t/\tau_{d,\lambda})$ with time scales $\tau_{d,\lambda}\approx 2$ days. We perform linear fits between $+4$ and $+10$ days to determine the post-peak decline rates (see Table~\ref{lc_paremeters}). The decline rate is wavelength dependent, being faster in the bluer bands ($\tau_{d,B}=1.73\pm0.05$ day or $0.63\pm0.02$\,mag day$^{-1}$) than the redder bands ($\tau_{d,i}=2.41\pm0.16$ days or $0.45\pm0.03$\,mag day$^{-1}$).
These decline rates are significantly faster than those of any previous rapidly evolving SN~I (see left panels of Figure~\ref{fig:lcs_compare} and $\S$~\ref{sec:comparison}). We have only sparse photometric coverage beyond $\sim +10$ days. Nevertheless, $i$-band images taken between $+16.3$ and $+40.2$ days suggest the decline-rate became substantially slower at $\approx 0.07$\,mag day$^{-1}$. Similar change in decline rate is seen in the late-time $i$-band LC of 2005ek. There are a few dozen fast transients discovered in large photometric surveys \citep{Drout2014, Pursiainen2018}, most of which do not have spectroscopic data for classification and generally have relatively sparse light-curve coverages. We compare 2019bkc with those objects in Appendix \S~\ref{sec:drout}.

We place an upper limit on the equivalent width (EW) of  \ion{Na}{1}D absorption at $z=0.0209$ of $<0.4$\,\AA, which corresponds to $E(B-V)_{\rm host}<0.07$ mag \citep{Phillips2013}. Thus we assume no host reddening. Note that we have not performed an exhaustive search for possible \ion{Na}{1}D absorption lines at redshift other than the adopted value. Taking into account the Galactic extinctions of $E(B-V)_{\rm Gal} = 0.06$ mag \citep{Schlafly2011} and $R_V = 3.1$, we compute the absolute peak magnitudes, assuming cluster membership, given in  Table~\ref{lc_paremeters}.  At $M^{\rm peak}_{r} = -17.29\pm0.10$\,mag, SN~2019bkc is comparable to an average SN~Ib/c \citep[see, e.g.,][]{Taddia2018} and also the rapidly declining  SN~2005ek \citep{Drout2013} and SN~2010X \citep{Kasliwal2010}, as shown in the right panel of Figure~\ref{fig:lcs_compare} and discussed below.

We construct a pseudo-bolometric LC by performing blackbody (BB) fitting (see Appendix~\ref{sec:bb}). At peak, SN~2019bkc reached $T_{\rm BB} \approx 8,900$~K and $L_{\rm BB} \approx 3 \times10^{42}$ erg s$^{-1}$. At late time (beyond $\sim$16 days) when only Sloan Digital Sky Survey (SDSS)-$i$ band photometry is available, the pseudo-bolometric luminosities are estimated by assuming the BB temperature stays constant at $T_{\rm BB}=4000$\,K.

\begin{table*}[t]
\caption{Photometric Parameters of SN~2019bkc}
\begin{center}
\small 
\begin{tabular}{cccccc}
	\hline
	Filter & 
	JD$^{\rm peak}$&
	$m^{\rm peak}_{\lambda}$  &  
	$M^{\rm peak}_{\lambda}$  &  
	$\Delta m_{10}(\lambda)^{a}$ & 
	Decline Rate$^b$   \\ 
	& & (mag) & (mag) & (mag) &  (mag day$^{-1}$)\\
	\hline
	$B$ & $2458547.39\pm0.04$&$17.77\pm0.02$&$-17.14\pm0.10$&$5.24\pm0.07$&0.63$\pm$0.02\\
	$V$ & $2458547.66\pm0.07$&$17.55\pm0.02$&$-17.30\pm0.10$&$4.16\pm0.07$&0.57$\pm$0.01\\
	$g$ & $2458547.22\pm0.08$&$17.64\pm0.03$&$-17.25\pm0.10$&$4.49\pm0.12$&0.56$\pm$0.02\\ 
	$r$  & $2458547.66\pm0.16$&$17.54\pm0.03$&$-17.29\pm0.10$&$3.85\pm0.10$&0.56$\pm$0.01\\ 
	$i$  & $2458547.57\pm0.14$&$17.57\pm0.03$&$-17.25\pm0.10$&$3.37\pm0.14$&0.45$\pm$0.03\\
	\hline
	\hline
 \end{tabular}
\end{center}
Notes.
\newline
$^a$ The difference in magnitude between peak and $+10$ days. 
\newline
$^b$ As measured by a linear fit between $+4$ and $+10$ days.
\label{lc_paremeters}
\end{table*}

\subsection{Optical Spectra}
\label{spectra}

\begin{figure}
\centerline{\includegraphics[width=10cm]{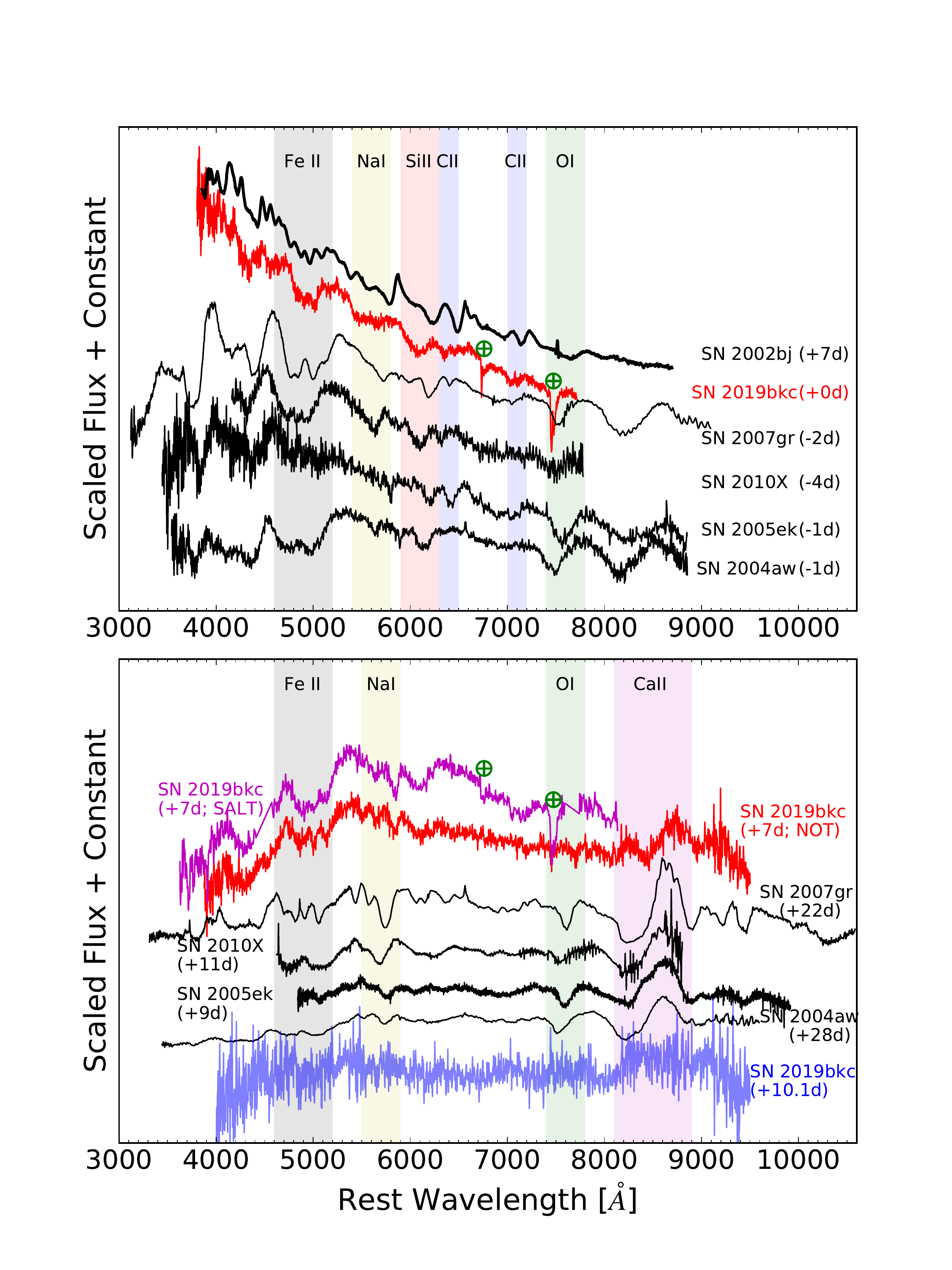}}
\caption{Comparison of SN 2019bkc with fast-declining SNe (SN 2002bj, SN 2005ek, SN 2010X) and other well-studied SNe Ic  (SN 2004aw, SN 2007gr). The wavelength regions of prominent lines commonly seen in SNe Ic near respective epochs are marked for guidance. The phases measured from peak (except for 2002bj, which is measured from the discovery date) are indicated in the parentheses. Top panel: spectra near maximum light. The regions around \ion{Na}{1} $\lambda$5890/5896, \ion{Si}{2} $\lambda$6355, \ion{C}{2} $\lambda$6580, \ion{C}{2} $\lambda$7234 and \ion{O}{1} $\lambda$7774 are marked with shaded regions. Bottom panel: post-peak spectra. The regions around \ion{Na}{1} $\lambda$5890/5896, \ion{O}{1} $\lambda$7774 and \ion{Ca}{2} triplet are marked with shaded regions. The absorption feature near 4600--5200 \AA which are usually attributed to iron-peak elements in SNe Ia are also marked in both panels. }
\label{fig:spectra_compare}
\end{figure}

Our spectra (Figure~\ref{fig:spectra_compare}) reveal a rapid evolution over 10 days.  At peak the spectrum consists of a blue continuum with a number of  broad absorption  features. A week later the continuum is much redder, consistent with $T_{\rm BB}$ dropping from $\sim$8,900 to $\sim$4000\,K, and shows substantial changes.
  
The spectral features of SN~2019bkc and their evolution do not have a good one-to-one match with any existing SN. We do not detect any hydrogen Balmer lines with P-cygni profiles seen in SNe II, so SN~2019bkc formally belong to Type~I. There are similarities in multiple features with the rapidly declining Type~I  SN~2010X and SN~2005ek, as well as with some SNe Ic, such as SN~2007gr \citep{Valenti2008} and SN~2004aw \citep{Taubenberger2006} aside from some relative velocity shifts (see Figure~\ref{fig:spectra_compare}). We use these similarities to aid line identifications. Given the spectral peculiarities, we do not regard all of these identifications as definitive, and we discuss the main lines and the velocities with relatively secure identifications below. Note that redshift $z=0.0209$ is assumed in the analysis, and the results and the associated velocities can change if the SN is at a different redshift. The commonly seen lines in SNe Ic at the respective epochs are marked for guidance in Figure~\ref{fig:spectra_compare}. More detailed discussion on the line comparisons and identifications are given in Appendix~\ref{sec:spec_features}.
  
In the first spectrum of SN~2019bkc, we identify several prominent absorption features (top panel of Figure~\ref{fig:spectra_compare}). The velocities measured from  \ion{Fe}{2} $\lambda$5169 and \ion{Si}{2} $\lambda$6355 absorption lines are $\sim$-9000 km~s$^{-1}$ and $\sim$-13000 km~s$^{-1}$ respectively. There are features near \ion{C}{2} $\lambda$6580 and \ion{C}{2} $\lambda$7234, but if they are both due to \ion{C}{2}, they have discrepant velocities differing by $\sim$2000 km~s$^{-1}$, so we do not regard the line identification of \ion{C}{2} as secure. There is an absorption feature near \ion{Na}{1} $\lambda$5890/5896, but it is much broader than the \ion{Na}{1} seen in other SNe. If \ion{Na}{1} is present in SN~2019bkc, it is likely severely blended with other lines, so we do not have a clear identification of \ion{Na}{1}. Strong \ion{O}{1} is commonly seen in normal Ibc, and it is also seen in SN~2005ek and 2010X, which is used as evidence supporting their progenitors are massive stars \citep[see, e.g.][]{Drout2013}. There is no strong \ion{O}{1} detected in our SN~2019bkc spectra.    
 
A week past peak, the \ion{Si}{2} feature appear to disappear, while the \ion{Fe}{2} lines remain visible. The \ion{Fe}{2} absorption velocity decreases quickly from $\sim$-9000\,km~s$^{-1}$ in the near-peak spectrum to $\sim$-2,500 km\,s$^{-1}$ one week after peak. Furthermore, some relatively narrow features around 5500\,\AA\,emerge (yellow shaded region in bottom panel of Figure~\ref{fig:spectra_compare}),  and the one at 5800\,\AA\,is likely \ion{Na}{1} $\lambda$5890/5896 with velocity of $\sim$-2900 km\,s$^{-1}$. The broad feature at  8500\,\AA\ is likely due to \ion{Ca}{2} NIR triplet. In our spectrum taken at $+10.1$ days, there is a tentative detection of \ion{Ca}{2} NIR at low signal-to-noise ratio (S/N), and if it is real, it has a large blueshift of $10000$\,km~s$^{-1}$.

\section{Discussion}
\label{discussion}
 
 \subsection{Comparison with Known Objects}
 \label{sec:comparison}
 
In Figure~\ref{fig:lcs_compare}, we compare the LCs of SN~2019bkc with several other rapidly declining SNe I with good photometric \textit{and} spectroscopic coverage, including SN~2002bj \citep{Poznanski2010}, SN~2005ek \citep{Drout2013}, SN~2010X \citep{Kasliwal2010}. We also include the LCs of  Type~Ic~SN~2007gr \citep{Valenti2008}, which we use extensively for spectroscopic comparisons in this work, and the fast-declining 1991bg-like Type Ia SN 1999by \citep{Silverman2012} to compare with a SN I not falling into the Ibc category.  In addition, we compare it with the fast transient KSN~2015K \citep{Rest2018} as it has an exquisite LC observed by {\it Kepler} and electromagnetic counterpart of merger of double NS, the ``kilonova'' GW 170817 \citep[also known as AT 2017gfo;][]{Villar2017}, which has extremely rapidly declining LCs.

SN 2019bkc declines faster than all of the SNe in the comparison sample except the kilonova. The decline between peak and +10 days  for SN~2019bkc  in SDSS-$r$ band is $\Delta$m$_{10}(r) = 3.85\pm0.10$\,mag (see Table~\ref{lc_paremeters} for $\Delta$m$_{10}$ in other bands; peak magnitudes from third-order polynomials fit and magnitudes at +10 days from linear extrapolation of data within +4 days $-$ +10 days are used in calculation of $\Delta$m$_{10}$), compared to $\Delta$m$_{10}(R) = 1.1\pm0.2$\,mag for SN~2002bj,  $\Delta$m$_{10}(R) = 1.85\pm0.15$\,mag for SN~2005ek,  $\Delta$m$_{10}(r) = 1.3\pm0.2$\,mag  for SN~2010X, and  $\Delta$m$_{10}(\rm Kepmag) = 2.10\pm0.03$\,mag for KSN~2015K. 

The right panel of Figure~\ref{fig:lcs_compare} shows the absolute $r/R$-band magnitude evolution. SN~2005ek, SN~2010X, and SN~2019bkc  are the most similar, with peak absolute magnitudes $\approx -$17~mag.  The remainder of the comparison sample exhibits significant luminosity and/or luminosity evolution differences compared to SN~2019bkc  at nearly all phases. In particular, SN~2002bj and KSN~2015K are substantially more luminous (by $\gtrsim$ 1.5 mag) at peak. It is worth noting that the reported luminosity of SN~2019bkc relies upon its redshift estimate $z=0.0209$ based on the likely association with galaxy cluster MKW1, and to reach similar peak luminosities with SN~2002bj and KSN~2015K would require that the SN be about two times further away.  Kilonova GW 170817, which does not share spectroscopic similarities with SN~2019bkc, is $\sim$1~mag dimmer than SN~2019bkc at peak, and it declines faster with $\Delta$m$_{10}(r)=5.0\pm0.3$\,mag. 

Among the comparison sample above, spectra of SN 2019bkc share some similarities with those of SN 2005ek, SN 2010X and SN 2007gr in which common features attributed to \ion{Fe}{2}, \ion{Na}{1}, \ion{Si}{2}, and \ion{Ca}{2} are seen (see Figure~\ref{fig:spectra_compare}). No clear signature of \ion{O}{1} $\lambda7773$ is seen in spectra of SN 2019bkc or comparatively weak if it is detected at all, and even for common species, SN 2019bkc's line profiles are less pronounced than the comparison objects, for example the iron peak absorption in near-peak spectrum and \ion{Ca}{2} NIR triplet in +7 days spectrum. In addition, there are significant differences in the  \ion{Si}{2} absorption velocities among these SNe (see the red-shaded region in the top panel of Figure~\ref{fig:spectra_compare}).

 \subsection{Interpretations}
 \label{sec:interpretations}

In this section, we discuss the possible energy sources and progenitor scenarios for SN~2019bkc. 

The rapid LC evolution implies a low ejecta mass. For photon diffusion out of an expanding spherical ejecta with velocity $v = 10^4 v_4\,{\rm km\,s^{-1}}$, the optical depth is of the order $\tau \sim c/v$ when the diffusion time $t_{\rm diff} = 4 t_{\rm d4}$ days equals the time since explosion. The ejecta mass is approximately $M \sim 4\,\pi\,R^2/ \kappa \tau \sim 0.08\,M_\odot v_4 t^2_{\rm d4} \kappa_{0.3}^{-1}$, where $R=v t_{\rm diff}$ and $\kappa = 0.3\kappa_{0.3}$ cm$^2$ g$^{-1}$ is the ejecta opacity. Given light-curve evolution timescales above half maxima rising to the peak and declining from the peak of $t_{1/2,\rm{rise}}\approx5$ days and $t_{1/2,\rm{decline}}\approx2$ days, respectively, for SN~2019bkc, it likely has a small ejecta mass $\sim$0.1\,M$_\odot$.

If the radiated energy is entirely from shock deposition \citep[see, e.g.][]{Waxman2017}, $E_{\rm rad}\sim Mv^2R_*/(2 v t_{\rm diff})\sim Lt_{\rm diff}$ and thus the progenitor radius $R_* \sim 5\times10^{12}$\,cm,  implying a massive star progenitor. 

The remote location of SN~2019bkc implies an isolated progenitor (or in an undetected globular cluster (GC)) in the outskirts of a galaxy halo or intracluster light, and it is thus probably associated with an old stellar population. Assuming that the supernova is at $z=0.0209$, the absence of a star-forming host galaxy down to $M_g\gtrsim-8.5$ and the non-detection of H$\alpha$  ($\lesssim\!10^{37}$~erg~s$^{-1}$) in our spectra make a massive star progenitor relatively unfavorable. However, we cannot completely rule out the possibilities of an extremely low-luminosity dwarf-galaxy host, association with an isolated star forming region forming stars at $\lesssim10^{-5} M_\odot\ {\rm yr^{-1}}$ or a hypervelocity runway massive star progenitor from a nearby member of the MKW1 cluster (see   Appendix~\ref{sec:environment}).

Next we consider  whether the LCs can be solely $^{56}$Ni-powered. We apply the integral relation of \citet{Katz2013} to the pseudo-bolometric LCs with low-order polynomial extrapolation to the explosion time ($t_{\rm exp}$). Significant $\gamma$-ray escape from low-density ejecta can produce a fast decline, and we freely vary the $\gamma$-ray escape timescale $t_0$ \citep[see][and references therein]{Stritzinger2006}. We  considered a broad range for $t_{\rm exp}$, but find no satisfactory match with data (see Appendix~\ref{sec:ni56} for a detailed demonstration on this). Therefore, energy deposition is unlikely to be dominated by $^{56}$Ni decay. The late-time $i$-band LC is consistent with decay of $^{56}$Ni of $\sim$0.001--0.01\,{\rm M}$_\odot$, subject to the uncertainty of $\gamma$-ray escape timescale.

\begin{figure}
\centerline{\includegraphics[width=10cm]{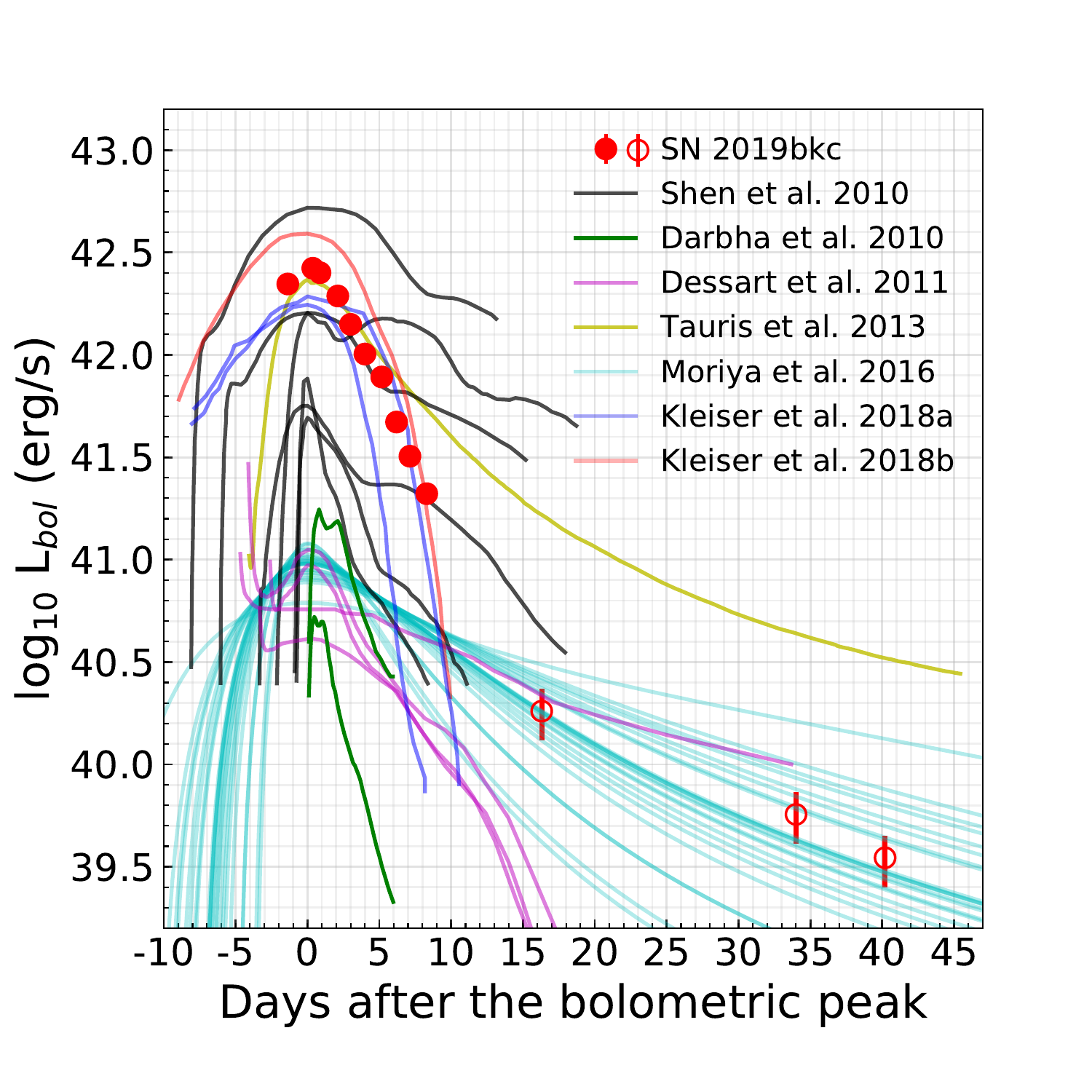}}
\caption{Pseudo-bolometric LC of SN~2019bkc, compared with synthetic LCs from various models. The open circles are estimated based on $i$-band data for $>15$ days. 
The models include: He-shell detonations (.Ia SNe; \citealt{Shen2010}), AIC\citep{Darbha2010}, an ultra-stripped ccSN producing little $^{56}$Ni \citep{Tauris2013a}, ccSNe of Wolf--Rayet stars producing no $^{56}$Ni  \citep{Dessart2011}, $^{56}$Ni-free stripped ccSNe  with circumstellar shells/envelop \citep{Kleiser2018a, Kleiser2018b} and various ECSNe  \citep{Moriya2016}. See Appendix~\ref{sec:comparemodels} for descriptions of specific models.}
\label{fig:Lbol_compare}
\end{figure}

Next we discuss various models interpreting rapid declining SNe in the literature. The first class of progenitor systems includes at least one low-mass compact star (e.g., a WD). In the .Ia scenario, a WD detonates from an accreted He layer \citep{Bildsten2007}, resulting in the production of  $\lesssim 0.02$ $M_{\odot}$ of  short-lived isotopes (e.g., $^{48}$Cr, $^{52}$Fe, $^{56}$Ni) and the ejection of $\sim$0.2--0.4 M$_{\odot}$ of material \citep{Shen2010}. Alternative WD-related scenarios that produce short-lived transients include AIC of O/Ne/Mg WDs  \citep[e.g.,][]{Dessart2006,Darbha2010}. Furthermore, several studies consider evolved massive stars as progenitors of rapidly declining SNe.
 \citet{Tauris2013a} modeled ultra-stripped ($\sim 1.5$~$M_{\odot}$) He-stars that explode following  Fe-core collapse and reproduce the LC of SN~2005ek. \citet{Moriya2010} studied dim ccSNe with fallback, in which a massive star ($\sim$13 $M_{\odot}$)  explodes with such low kinetic energy  that only a small amount of material (including $^{56}$Ni) is ejected and  the majority remains bound to a compact remnant.  \citet{Kleiser2018a} proposed explosion of $\sim$2--4 $M_{\odot}$ He-stars with an extended envelope ($\gtrsim$ 25 $R_{\odot}$) with no $^{56}$Ni ejected to explain SN~2010X-like SNe, and the LC is powered by thermal energy deposited within the ejecta by the explosion shock wave following the core bounce \citep{Dessart2011,Kleiser2018b}. Another possibility is electron capture SNe (ECSNe) of $\approx$ 8~$M_{\odot}$ AGB stars \citep[e.g.,][]{Pumo2009,Moriya2016},  and in that case, due to the lack of CSM lines, the progenitor in a binary system would be necessary.  
 
The representative LCs calculated from the fast-evolving models mentioned above are compared to pseudo-bolometric LCs of SN~2019bkc and shown in Figure~\ref{fig:Lbol_compare}. The powering mechanisms of these models fall into two categories -- either shock deposition or $^{56}$Ni decay -- to explain the early-phase LCs near the peak. For the models whose radiation energy purely come from shock deposition, they cannot reproduce the late-time $i-$band LC. The late-time $i-$band LC can be powered by $\sim 0.001-0.01\,M_{\odot}$ of $^{56}$Ni, which is insufficient to power the early-time LCs.  \cite{Kleiser2018a,Kleiser2018b} proposed a class of models in which the LCs are predominantly powered by the shock cooling produced from massive, hydrogen-free stars, and such models can also produce a small amount of $^{56}$Ni powering a late-time tail in the LC. The models by \cite{Kleiser2018a,Kleiser2018b} may explain the overall LCs of SN~2019bkc, however, their massive star progenitors are in tension with the likely old stellar environment. More detailed discussion on the models are given in Appendix~\ref{sec:comparemodels}. Finally, we note a caveat that the comparisons presented here depend on the adopted $z=0.0209$ redshift.

\section{Summary}
\label{sec:summary}
In this section, we summarize the key properties of SN 2019bkc.
 \begin{itemize}
    \item  SN~2019bkc is the most rapidly declining SN I (i.e., no hydrogen in the spectra) yet reported with a post-peak decline rate of $\Delta{m_{10}(r)}=3.85\pm0.10$ mag. SN 2019bkc has a significantly slower rise relative to the decline around the peak, which reaches bolometric luminosity $L_{bol}^{peak}=2.9\pm0.4\times10^{42}$ erg s$^{-1}$, adopting a redshift $z=0.0209$. The late-time LC shows a much slower decline, which is consistent with decay of $^{56}$Ni at $\sim$0.001--0.01\,$M_{\odot}$, but the overall LC cannot be powered predominantly by $^{56}$Ni decay.
     
    \item There is no evidence for the presence of hydrogen in spectra of SN~2019bkc. The spectra of SN 2019bkc do not resemble those of fast-declining but more luminous SN 2002bj and share some similarities with spectra of fast-evolving SN~2005ek and SN~2010X with comparable peak luminosities if adopting a redshift $z=0.0209$.
    
    \item SN~2019bkc is ``hostless'' with  no apparent host galaxy detected down to $\sim -8.5$ mag adopting a redshift $z=0.0209$ and likely exploded in an intracluster environment. This suggests a link with an older stellar  population and disfavors a massive star progenitor. SN~2019bkc suggests being cautious about assuming that fast-declining ``hostless'' transients are Galactic, cataclysmic variable outbursts as is common in traditional SN classification programs.
    
     \item The combination of light-curve properties and likely association with old stellar environment poses challenge to explain SN~2019bkc with existing models for fast-evolving SNe.

\end{itemize}

\acknowledgments
We thank Boaz Katz, Takashi Moriya, and Thomas Tauris for their help. P.C., S.D.,\ and S.B.\ acknowledge NSFC 11573003. We acknowledge Telescope Access Program (TAP) funded by the NAOC, CAS, and the Special Fund for Astronomy from the Ministry of Finance. We acknowledge SUPA2019A-002 (PI: M.D. Stritzinger) via OPTICON. M.S. and S.H. are supported by a project grant (8021-00170B) from the IRF, Denmark and a grant (13261) from VILLUM FONDEN.  J.S. acknowledges support from the Packard Foundation. C.S.K. is supported by NSF grants AST-1515876, AST-1515927 and AST-1814440 as well as a fellowship from the Radcliffe Institute for Advanced Study at Harvard University. E.W.P.\ acknowledges NSFC.\ 11573002.  E.A. acknowledges NSF grant AST-1751874. D.A.H.B. acknowledges South African National Research Foundation. M.G. is supported by the Polish NCN MAESTRO grant 2014/14/A/ST9/00121. K.D.F. is supported by Hubble Fellowship grant HST-HF2-51391.001-A.  S.B. is partially supported by the PRIN- INAF 2016  (P.I. M. Giroletti) S.B. and A.A.S acknowledge NSF grant AST-1814719. N.E.R. acknowledges Spanish MICINN grant ESP2017-82674-R and FEDER funds.

This work is partly based on NUTS observations made with NOT, operated by the NOT Scientific Association at the Observatorio del Roque de los Muchachos, La Palma, Spain, of IAC.  ALFOSC is provided by IAA under a joint agreement with the University of Copenhagen and NOTSA. NUTS is funded in part by IDA.  Some This Letter includes data via \textit{Magellan} Telescopes at LCO, Chile. This work is partly based on observations with SOAR, which is a joint project of the Minist\'{e}rio da Ci\^{e}ncia, Tecnologia, Inova\c{c}\~{o}es e Comunica\c{c}\~{o}es (MCTIC) do Brasil, NOAO, UNC, and MSU.
Some observations were obtained with SALT under the Large Science Programme on transients, 2018-2-LSP-001 (PI: DAHB). Polish support of this SALT program is funded by grant No. MNiSW DIR/WK/2016/07.

\bibliographystyle{apj}
\bibliography{ms}

\appendix

\begin{figure*}[htbp]
\centerline{\includegraphics[width=14cm]{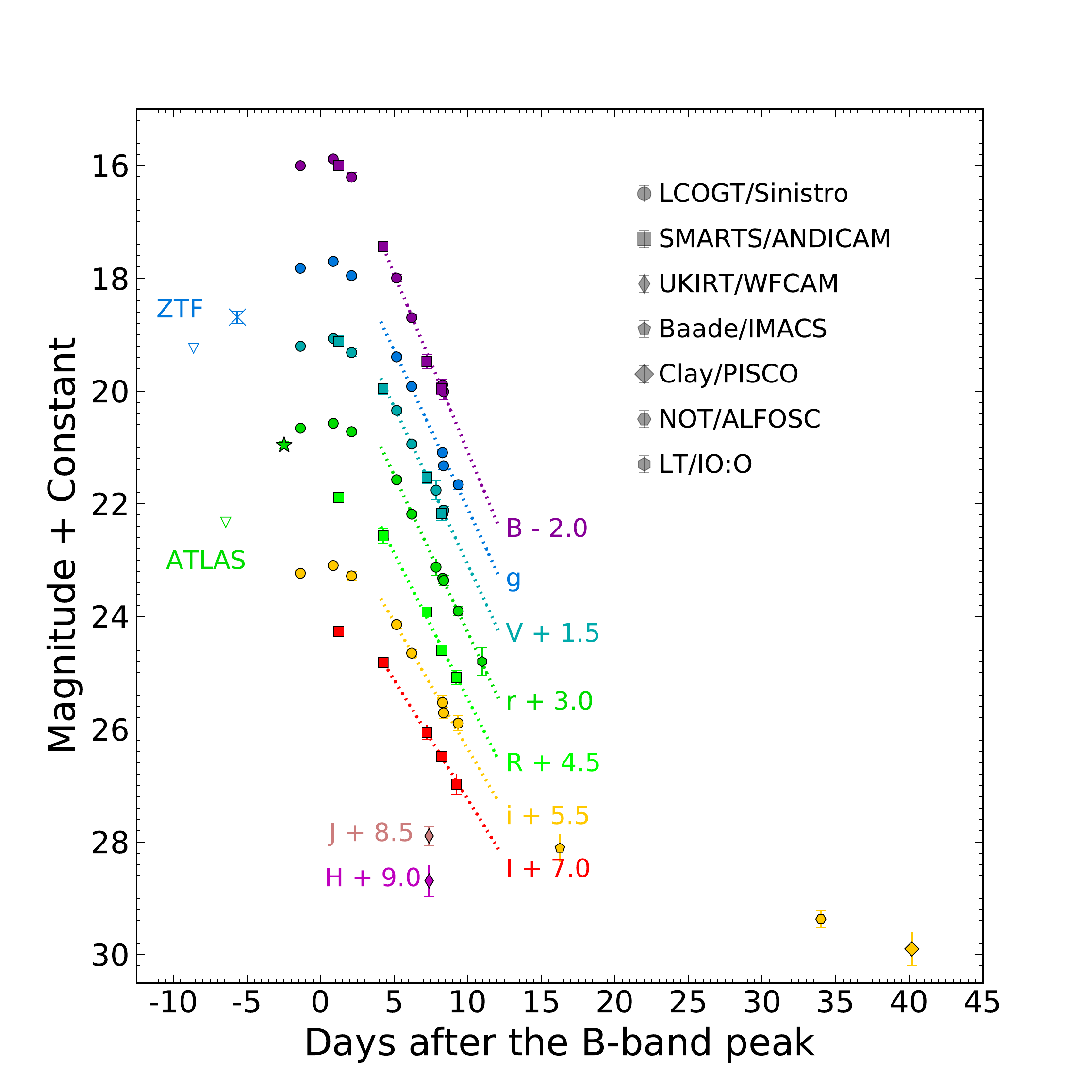}}
\caption{Multi-band LCs of SN~2019bkc. Dotted lines are linear fits to the data points between $+4$ and $+10$ days. The upper limits from ATLAS and ZTF are shown with open down-pointing triangles.}
\label{fig:sn19bkc_lcs}
\end{figure*}

\section{Photometry}

\subsection{Photometric Observation}
\label{sec:phot_obs}

Our optical and NIR images for SN 2019bkc were taken at the 1 m LCOGT, the SMARTS, UKIRT, NOT, LT, and the 6.5 m \textit{Magellan} telescopes.   

The majority of our optical images of SN 2019bkc were obtained with the LCOGT 1 m telescopes at Siding Spring Observatory (SSO), South African Astronomical Observatory (SAAO), and Cerro Tololo Interamerican Observatory (CTIO). Optical and NIR images were taken with A Novel Dual Imaging CAMera (ANDICAM) on SMARTS. The source is not reliably detected on any of the SMARTS NIR images, so they are excluded from further photometry. We obtained a single epoch of $JH$-band images with the WFCAM on UKIRT. All optical and NIR science images except the epoch obtained with Clay telescope (see below) were reduced following standard procedures including  bias/dark-frame and flat-field corrections, sky-subtraction, and the combination of multiple, dithered exposures.  

On the night of 2019 Apr 13 UT, using the Parallel Imager for Southern Cosmological Observations\citep[PISCO;][]{Stalder14} on the 6.5 m \textit{Magellan}/Clay telescope, we acquired ten $120\textrm{s}$ simultaneous $\{g, r, i, z\}$ exposures of SN 2019bkc. We reduced these data using the standard PISCO Rapid Analysis Toolkit (Brownsberger19, in prep), and we briefly summarize this reduction procedure here.  PISCO consists of four charge-coupled devices (CCDs; one per band), and each CCD has two amplifiers.  We apply an overscan correction to every amplifier for every image by subtracting from each amplifier a linear fit to the overscan pixels acquired as part of every readout.  We  median combine a stack of the 18 overscan-corrected bias exposures acquired that night, and subtract the resulting `master bias' from the flat and science images.  We acquired 49 twilight flat images during our observations.  However, not every flat can be used in the correction of every single-band CCD.  We arrange the twilight flat image into four stacks, one for each band.  A particular flat field is added to the stack for a particular band if and only if it has a characteristic surface brightness of between 5k and 15k ADU in that band, averaged between the two amplifiers.  One flat can be placed into zero, one, or multiple single-band stacks.  We normalize each overscan and bias corrected flat so that its median value is unity for the appropriate CCD, and then median combine each stack of normalized flats.  We thus acquire a separate 'master flat' field for each of the four PISCO filters.  We overscan and bias correct each of the 10 PISCO observations of SN 2019bkc, and separate each exposure into four single-band science images.  We divide each of these 40 single-band images by the appropriate master-flat.  We use the plate-solution algorithm provided astrometry.net \citep{Lang10} to determine the World Coordinate System (WCS) solution for each image.  We use this WCS solution and the SWarp coaddition software \citep{Bertin02} to produce our final PISCO $\{g,r,i,z\}$ images of SN 2019bkc.
 
Point-spread-function (PSF) photometry is computed with the DoPHOT \citep{Schechter1993} package for all images except for the PISCO images, for which we perform aperture photometry with \texttt{apphot} task of \texttt{IRAF}. The optical band photometry is calibrated relative to  APASS photometry (DR9; \citealt{Henden2016}) of stars in the field of SN 2019bkc, and the NIR photometry is calibrated relative to Two Micron All Sky Survey (2MASS) stars \citep{Skrutskie2006}. The APASS magnitudes in Johnson--Cousin $B$-, $V$- and Sloan $g$-, $r$-, and $i$- bands are directly used for photometry calibration of corresponding band, and are transformed to Johnson--Cousin $R$, $I$ magnitudes from SDSS $r$, $i$ magnitudes with $R = r-0.2936\times (r-i)-0.1439$ and $I=r-1.2444\times(r-i)-0.3820$ before applying the calibration. Template subtraction is not required to measure the SN's flux as there is no detectable host. All photometry for SN 2019bkc are reported in Table~\ref{tab:photometry} and the LCs are shown in Figure~\ref{fig:sn19bkc_lcs}.
 
 \begin{figure}[htbp]
\centerline{\includegraphics[width=16cm]{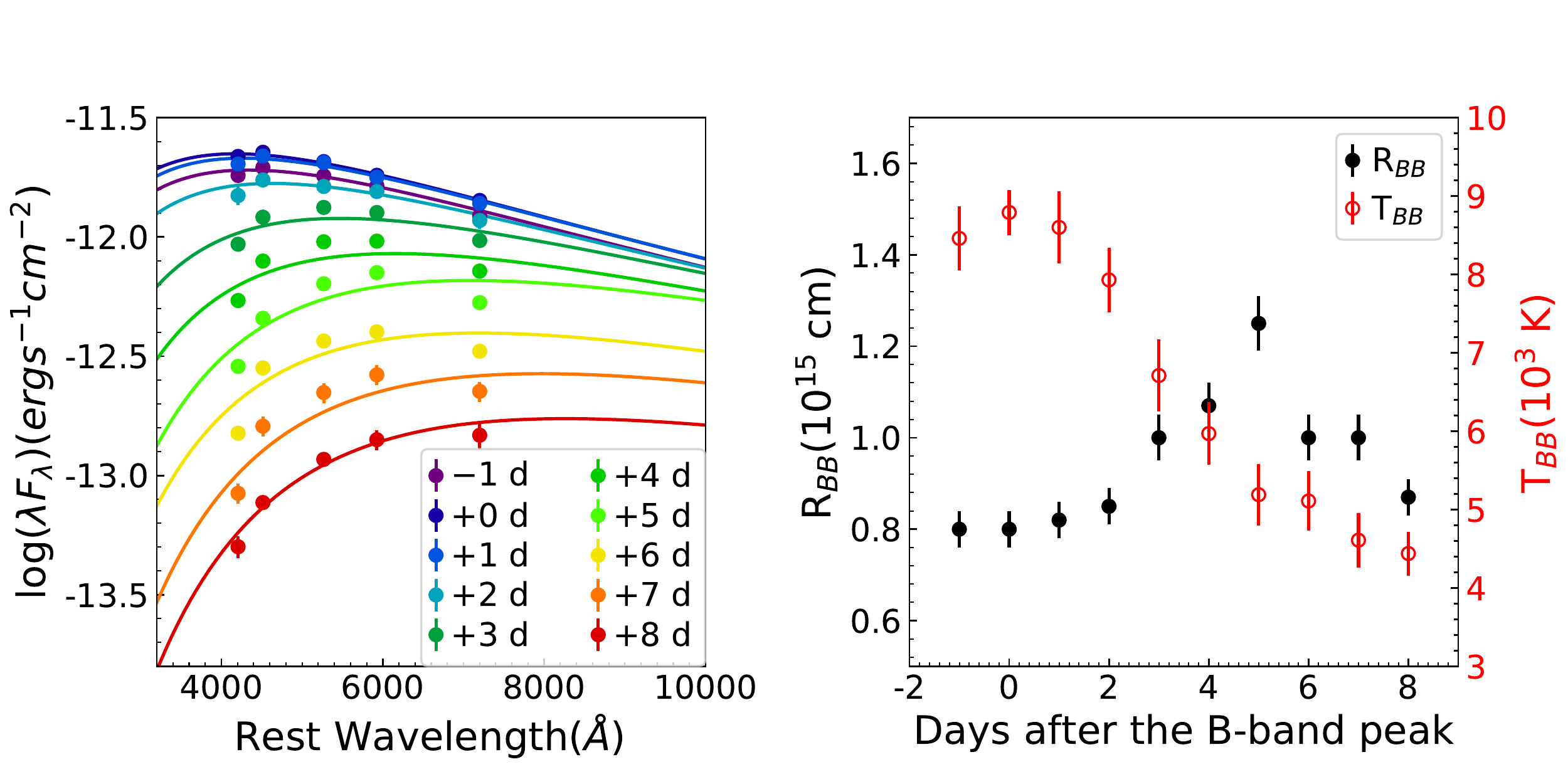}}
\caption{Left panel: SEDs constructed from our broad-band photometry. Over-plotted to each SED  is the line corresponding to the best-fit BB function. Right panel:  time evolution of best-fit BB parameters $T_{BB}$ and $R_{BB}$.}
\label{fig:bbfit_sed}
\end{figure}

\subsection{Spectral Energy Distribution (SED) and Bolometric LC}
 \label{sec:bb}
The optical SEDs shown in the left panel of Figure~\ref{fig:bbfit_sed} are modeled as blackbodies (solid lines), from which we get the effective temperature ($T_{\rm BB}$), and radius ($R_{BB}$) evolution of SN 2019bkc as shown in the right panel of Figure~\ref{fig:bbfit_sed}. The $R_{BB}$ evolution shows an initial increase from $\sim 0.8\times10^{15}$ cm up to $\sim 1.3\times10^{15}$ cm until $+$5 days, followed by a modest decline over the next several days, consistent with the typical radius evolution for the homologously expanding ejecta of a SN \citep[see, e.g.,][]{Liu2018} rather than some other type of transients such as a tidal disruption event. Following an initial increase as the transient reaches peak brightness, both the luminosity and temperature subsequently decline as the ejecta expand and cool. 
 
The bolometric LC (shown in Figure~\ref{fig:Lbol_compare}) for SN 2019bkc is constructed using the results derived from BB fitting at phases $<9$ days after $B$-band peak. SN 2019bkc has a peak bolometric luminosity of  $L_{bol}^{peak}=2.9\pm0.4\times10^{42}$ erg s$^{-1}$. We estimate the late-time bolometric luminosities at epoch of +16.3, +34.0, and 40.2 days after $B$-band peak with single SDSS-$i$ band photometry by assuming that the BB temperature stay constant at these epochs at  $T_{BB}=4000$ K. We add an extra 20\% uncertainties in quadrature for late-time bolometric LCs to account for possible systematic errors.  
 
 \begin{table*}
\scriptsize
\caption{Optical and NIR photometry of SN 2019bkc}
\begin{center}
\begin{tabular}{ccccccccccccc}
\hline
\hline
 Date       & JD $-$        & Phase & $B$         & $V$         & $R$        & $I$        & $g$          & $r$         & $i$         & $J$        & $H$        &Telescope\\
               & 2458000 & (days) &(mag)  &(mag)  &(mag)   &(mag)   &(mag)   &(mag) &(mag)  &(mag) &(mag)  & \\
 \hline 
2019 Mar 3 & 546.03 & $-$1.4& 18.00(03) & 17.71(03) & $\cdots$ & $\cdots$ & 17.82(02)  & 17.66(03) & 17.73(05) & $\cdots$ &$\cdots$ &LCOGT \\
2019 Mar 5 & 548.27 &  0.9  & 17.88(04) & 17.57(03) & $\cdots$ & $\cdots$ & 17.70(03)  & 17.57(03) & 17.60(04) & $\cdots$ & $\cdots$ &LCOGT \\
2019 Mar 6 & 548.64 &  1.2  & 18.00(05) & 17.62(10) & 17.39(09)& 17.26(08)&  $\cdots$  & $\cdots$  & $\cdots$  & $\cdots$& $\cdots$ &SMARTS \\  
2019 Mar 7 & 549.51 &  2.1  & 18.21(09) & 17.82(06) &  $\cdots$& $\cdots$ & 17.95(05)  & 17.72(05) & 17.78(08) & $\cdots$ & $\cdots$ &LCOGT\\
2019 Mar 9 & 551.65 &  4.3  & 19.44(07) & 18.46(10) & 18.07(13)& 17.81(07)&  $\cdots$  & $\cdots$  & $\cdots$  & $\cdots$ & $\cdots$ &SMARTS\\  
2019 Mar 10 & 552.57 &  5.2  & 20.00(06) & 18.84(05) &  $\cdots$& $\cdots$ & 19.40(04)  & 18.57(04) & 18.64(06) & $\cdots$ & $\cdots$ &LCOGT\\
2019 Mar 11 & 553.60 &  6.2  & 20.70(06) & 19.44(05) &  $\cdots$& $\cdots$ & 19.92(03)  & 19.19(05) & 19.15(06) & $\cdots$ & $\cdots$ &LCOGT\\
2019 Mar 12 & 554.63 &  7.2  & 21.48(12) & 20.04(10) & 19.42(08)& 19.05(13)&  $\cdots$  & $\cdots$  & $\cdots$  & $\cdots$ & $\cdots$ &SMARTS\\  
2019 Mar 12 & 554.78 &  7.4  & $\cdots$  & $\cdots$  &  $\cdots$& $\cdots$ &  $\cdots$  & $\cdots$  & $\cdots$  & 19.40(17)& 19.69(28)&UKIRT\\ 
2019 Mar 12 & 555.25 &  7.9  & $\cdots$  & 20.26(17) &  $\cdots$& $\cdots$ & 21.10(06)  & 20.13(15) & $\cdots$  & $\cdots$ & $\cdots$ &LCOGT\\  
2019 Mar 13 & 555.62 &  8.2  & 21.96(10) & 20.68(12) & 20.10(07)& 19.48(09)&  $\cdots$  & $\cdots$  & $\cdots$  & $\cdots$ & $\cdots$ &SMARTS\\  
2019 Mar 13 & 555.70 &  8.3  & 21.89(11) & 20.68(07) &  $\cdots$& $\cdots$ & 21.33(07)  & 20.32(10) & 20.03(13) & $\cdots$ & $\cdots$ &LCOGT\\
2019 Mar 13 & 555.77 &  8.4  & 22.01(14) & 20.62(07) &  $\cdots$& $\cdots$ & 21.66(08)  & 20.36(09) & 20.21(10) & $\cdots$ & $\cdots$ &LCOGT\\
2019 Mar 14 & 556.62 &  9.2  &  $\cdots$ & $\cdots$  & 20.58(12)& 19.98(19)&  $\cdots$  & $\cdots$  & $\cdots$  & $\cdots$ & $\cdots$ &SMARTS\\  
2019 Mar 14 & 556.76 &  9.4  &  $\cdots$ & $\cdots$  &  $\cdots$& $\cdots$ &  $\cdots$  & 20.91(09) & 20.39(13) & $\cdots$ & $\cdots$ &LCOGT\\
2019 Mar 15 & 558.38 & 11.0  &  $\cdots$ & $\cdots$  &  $\cdots$& $\cdots$ &  $\cdots$  & 21.80(25) & $\cdots$  & $\cdots$ & $\cdots$ &LT\\
2019 Mar 21 & 563.67 & 16.3  &  $\cdots$ & $\cdots$  &  $\cdots$& $\cdots$ &  $\cdots$  & $\cdots$  & 22.61(25) & $\cdots$ & $\cdots$ &Baade\\
2019 Apr 7 & 581.40 & 34.0  &  $\cdots$ & $\cdots$  &  $\cdots$& $\cdots$ &  $\cdots$  & $\cdots$  & 23.87(15) & $\cdots$ & $\cdots$ &NOT\\
2019 Apr 14 & 587.58 & 40.2  &  $\cdots$ & $\cdots$  &  $\cdots$& $\cdots$ &  $\cdots$  & $\cdots$  & 24.40(30) & $\cdots$ & $\cdots$ &Clay\\
 
\hline
\end{tabular}
\end{center}
\label{tab:photometry}
\end{table*}

\section{Spectroscopy}

\subsection{Spectroscopic Observation}
\label{sec:spec_obs}

\begin{figure}[htbp]
\centerline{\includegraphics[width=16cm]{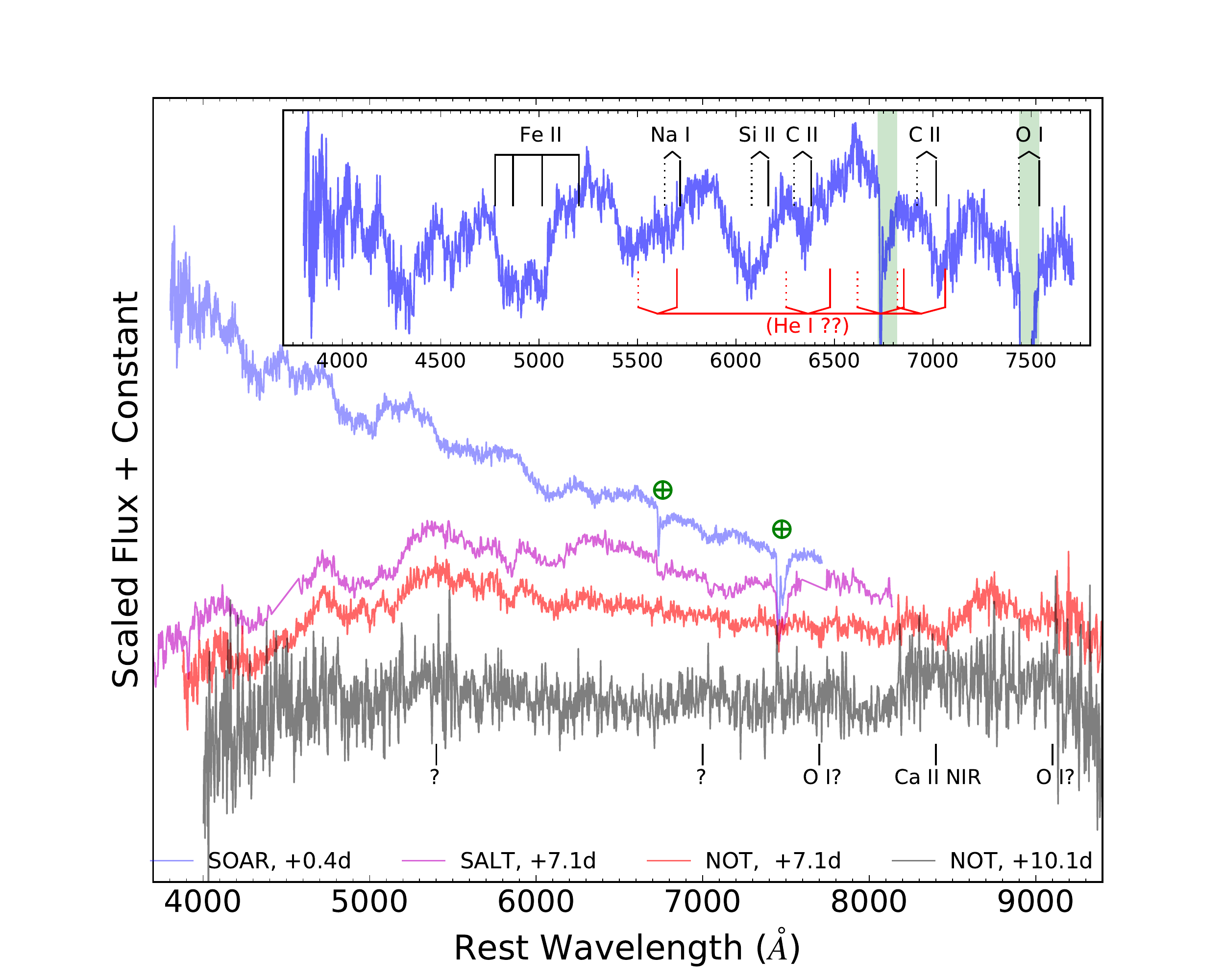}}
\caption{Optical spectra of SN 2019bkc. The SOAR spectrum is normalized and shown in inset panel. The regions affected by telluric lines are shaded and the shaded region width is 100\,\AA. Tentatively identified ions are labeled (black) along with the blue-shifted line centers where the solid lines are blue-shifted 9000 km s$^{-1}$ and dotted lines are blue-shifted 13000 km s$^{-1}$.  Line centers for strong persistent line of \ion{He}{1} are also indicated in red with solid lines (blue-shifted 9000 km s$^{-1}$) and dotted lines (blue-shifted 19000 km s$^{-1}$).}
\label{fig:spec_19bkc}
\end{figure}

We obtained four visual-wavelength spectra on +0.4 day (SOAR), +7.1 days (SALT), +7.1 days (NOT), and +10.1 days (NOT) relative to $B$-band peak. The spectroscopic data are reduced and calibrated following standard procedures (such as bias subtraction, flat-fielding, cosmic ray rejections, spectral extractions, and wavelength calibrations using arc lamp) using  \texttt{IRAF}. Each science spectrum was calibrated using a  spectrophotometric standard observation. Finally, each spectrum was color-matched to the broad-band colors of SN~2019bkc. The log of spectroscopic observations is given in Table~\ref{tab:speclog}. The full spectral sequence for SN 2019bkc is shown in Figure~\ref{fig:spec_19bkc} where the SOAR spectrum is normalized and shown in the top inset panel to highlight the absorption features. All reduced spectra are made available at the Weizmann Interactive Supernova Data Repository \citep{Yaron2012}.

\subsection{Notable Spectroscopic Features}
\label{sec:spec_features}

\begin{figure}[htbp]
\centerline{\includegraphics[width=18cm]{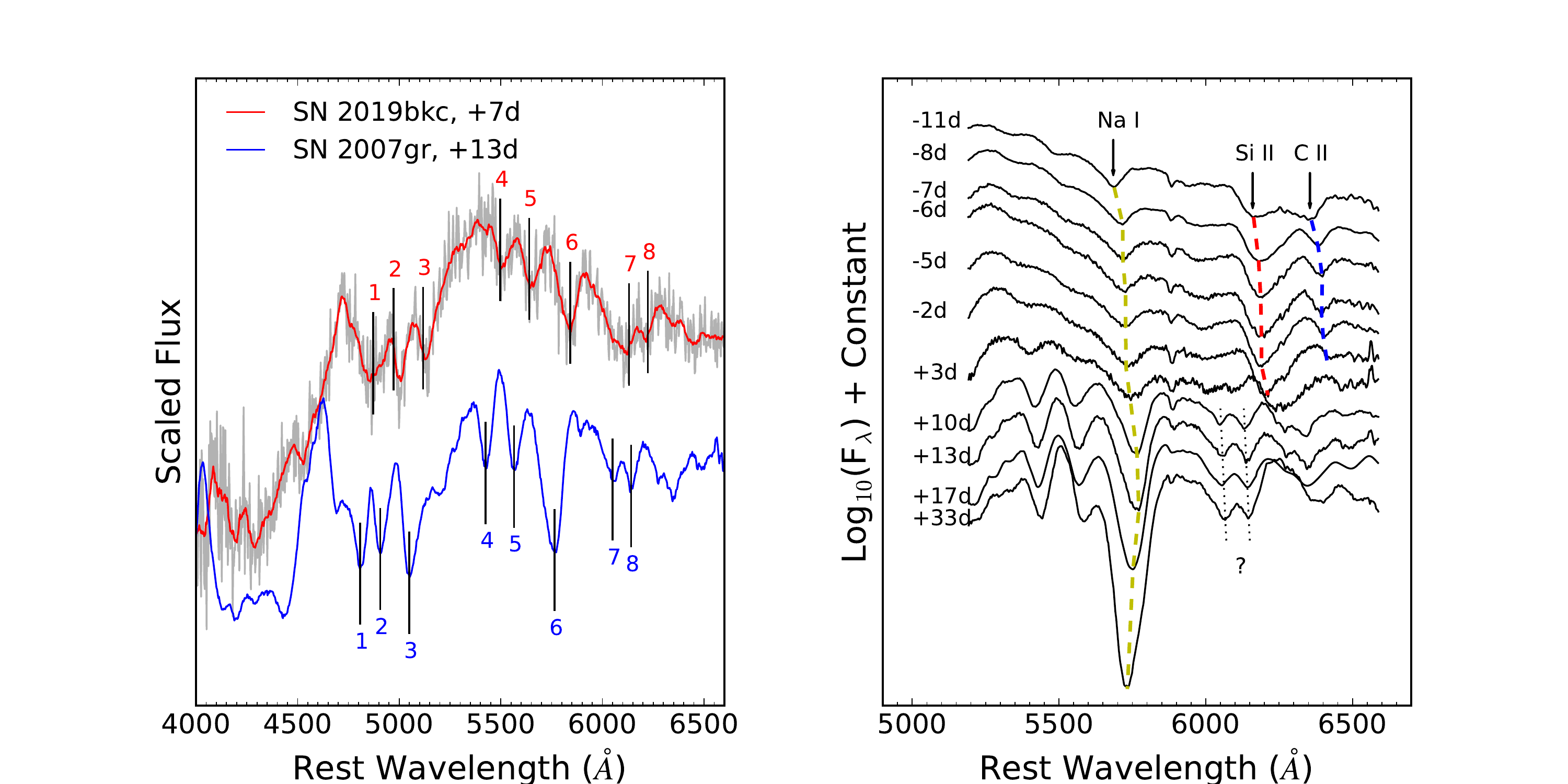}}
\caption{Left panel: comparison of SN 2019bkc +7 days spectrum and SN 2007gr +13 days spectrum in wavelength range of 4000--6500 \AA. Some conspicuous absorption features are marked and labeled for SN 2007gr (blue), which are simply red-shifted 4000 km s$^{-1}$ in SN 2019bkc (red). Right panel: spectral evolution of SN 2007gr from around +10 days before LC peak to around one month after that.  }
\label{fig:spec_features}
\end{figure}

In this section, we describe how we carry out line identification by making comparisons with well-studied SNe.
In Figure~\ref{fig:spectra_compare},  spectra of SN 2019bkc are compared with fast-declining SNe (SN 2002bj, SN 2005ek, SN 2010X) and other well-studied SNe Ic  (SN 2004aw, SN 2007gr) with near-peak  spectra in top panel and ``late''-time spectra in the bottom panel. The overall spectral features of 2019bkc are very different from the fast-declining SN 2002bj, which has clear He detection in the spectrum and nominally a SN Ib \citep{Poznanski2010}. In general, 2019bkc shares some similarities with SNe Ic but has a number of distinct differences, which are discussed below.

In Figure~\ref{fig:spectra_compare}, spectral lines commonly seen in SNe Ic are marked in color-shaded regions and labeled the responsible ions. From the top panel (i.e., near-peak spectra), we can identify some features that appear to be shared by  2019bkc and other SNe Ic in wavelength regions of 4600--5200\,\AA, 5900--6300\,\AA\,,  and 6300--6500\,\AA\, (see shaded regions in gray, red, and blue in Figure~\ref{fig:spectra_compare}), and they are mainly attributed to \ion{Fe}{2},  \ion{Si}{2}, and \ion{C}{2}, respectively \citep[see, e.g.,][]{Valenti2008,Drout2013}. Next we discuss the line identifications in these regions.

We measured the Doppler velocity of $\sim$-9000 km s$^{-1}$ from \ion{Fe}{2} $\lambda$5169 absorption line in near-peak spectrum of SN 2019bkc. At one week after peak, the \ion{Fe}{2} absorption features remain in the spectrum and the velocity decreases fast to $\sim$-2500 km s$^{-1}$. As shown in the left panel of Figure~\ref{fig:spec_features}, there are striking similarities in eight absorptions features between SN 2019bkc (+7 days) and SN 2007gr (+13 days) with a common relative velocity shift of 4000 km\,s$^{-1}$, implying that they have the similar compositions and physical conditions at the compared phases. Among them, features 1, 2, and 3 are identified as  \ion{Fe}{2} for SN 2007gr \citep{Valenti2008}. Feature 6 is identified as \ion{Na}{1} for SN 2007gr while features 4, 5, 7, and 8 had no ion identifications in \cite{Valenti2008}.

The velocities measured from \ion{Si}{2} $\lambda 6355$ absorption lines near the peak show a large diversity among comparison objects. SN 2019bkc and SN 2010X have the highest velocities $\sim$13000 km\,s$^{-1}$, SN 2005ek, and SN 2007gr have the lowest velocities of $\sim$7000 km\,s$^{-1}$, and SN 2004aw resides in the middle with velocity of $\sim$10000 km\,s$^{-1}$. The \ion{Si}{2} $\lambda$6355 feature of 2019bkc is broader and relatively strong compared  to \ion{Fe}{2},  which cautions that the feature centering around 6100\,\AA might be contaminated by other ions so that the velocity reported above for \ion{Si}{2} $\lambda$ 6355 might be affected.  In the +7 days spectra of 2019bkc, there is an absorption feature at $\sim$6000\,\AA. If it is attributed to \ion{Si}{2} $\lambda$6355, it translates to a blueshift velocity of $\sim$12000 km\,s$^{-1}$, which is inconsistent with velocity of $\sim$2500 km\,s$^{-1}$ from \ion{Fe}{2}. This line corresponds to feature 7 in Figure~\ref{fig:spec_features}, and its corresponding feature has no ion identification in \citep{Valenti2008}. In fact, as shown in the right panel of Figure~\ref{fig:spec_features}, the  \ion{Si}{2} and \ion{C}{2} lines both disappear at $\sim10+$ days. We conclude that \ion{Si}{2} $\lambda 6355$ (and  \ion{C}{2}) lines are not detected in $+7$ days spectrum of 2019bkc. In the near-peak spectrum, there are features near \ion{C}{2} $\lambda$6580 and \ion{C}{2} $\lambda$7234. However, if they are both due to \ion{C}{2}, the derived velocities differ by $\sim 2000$ km~s$^{-1}$. We do not regard their identifications as secure. As discussed above, we do not have secure identification of \ion{C}{2} in the $+7$ days spectrum either.

\ion{O}{1} is one of the most conspicuous feature in spectra of normal SNe Ib and SNe Ic.
Strong \ion{O}{1} lines were identified in 2005ek \citep{Drout2013} and 2010X \citep{Kasliwal2010}. \cite{Drout2013} found that the ejecta of SN 2005ek are dominated by oxygen ($\sim$86\%) by abundance modeling, and this was used to argue that its progenitor was a massive star. In near-peak spectra, the \ion{O}{1} $\lambda$7774 lines  (see the shaded region in green in Figure~\ref{fig:spectra_compare}) are stronger than or comparable to \ion{Si}{2} $\lambda$6355 lines in SN 2004aw, SN 2005ek, SN 2007gr, and SN 2010X.  \ion{O}{1} $\lambda$7774 at similar or greater strength compared to \ion{Si}{2} $\lambda$6355 is not detected in SN 2019bkc, while detecting significantly weaker  \ion{O}{1} $\lambda$7774 is hindered by telluric lines (see the light-green regions for the positions of telluric lines in inset panel of Figure~\ref{fig:spec_19bkc}). \ion{O}{1} $\lambda$7774 is not detected in +7 days spectrum of SN 2019bkc, while for comparison objects \ion{O}{1} $\lambda$7774 continues to be strong. We conclude that strong oxygen is likely not present for SN 2019bkc.

Next, we examine whether helium is present in the spectra of SN 2019bkc. There are degeneracies in locations between  \ion{He}{1} ($\lambda$5876, $\lambda$6678, $\lambda$7281), \ion{Na}{1} ($\lambda$ 5890/5893), and \ion{C}{2} ($\lambda$6580, $\lambda$7234), making \ion{He}{1} line detection for SNe Ibc challenging \citep[see, e.g.,][]{Matheson2001}. 
Helium was not detected in 2010X \citep{Kasliwal2010} or 2005ek \citep{Drout2013}, yet its presence cannot be conclusively ruled out in either case. With available data, we cannot ambiguously distinguish \ion{He}{1} $\lambda$5876 from \ion{Na}{1} $\lambda$5890/5896. The spectrum of SN 2019bkc has a distinct deep absorption feature at $\sim$5500\,\AA (near the left edge of yellow shaded region in the top panel of Figure~\ref{fig:spectra_compare}; see also the normalized spectrum in the inset panel of Figure~\ref{fig:spec_19bkc}), and there is no such feature in the spectra of any comparison objects. \cite{Clocchiatti1996} reported detection of high-velocity ($v \sim$ 16900 km s$^{-1}$) helium features in optical spectra of SN 1994I, which implies high-velocity helium in the outer portion of the ejecta. Here we cannot rule out the possibility that the above-mentioned absorption feature at $\sim$5500\,\AA\, is from high-velocity ($v\sim$19000 km s$^{-1}$) helium in the ejecta. 

Our last spectrum (NOT, +10.1d) for SN 2019bkc is of low S/N while there appear to be some emission features. The tentative detections of emission features are marked in Figure~\ref{fig:spec_19bkc} along with some possible identification. The tentative detection of  \ion{Ca}{2} NIR feature has a significant blueshift at $\sim$10000\,km\,s$^{-1}$, and this would imply an increase in velocity from $\sim 3$ days before.

\begin{table*}[htbp]
\caption{Summary of Spectroscopic Observations of SN 2019bkc}
\begin{center}
\begin{tabular}{ccccc}
\hline
\hline
 Date      &   JD   &  Phase(day) &    Range(\AA) &   Telescope   Instrument  \\
 \hline
2019 Mar 5 & 2458547.77  & 0.4  &   3900--7900   &  SOAR/GTHS$^a$  \\
2019 Mar 11 & 2458554.45  & 7.1  &   3700--8300   &  SALT/RSS$^b$  \\
2019 Mar 11 & 2458554.53  & 7.1  &   4000--9700   &  NOT/ALFOSC$^c$\\
2019 Mar 14 & 2458557.47  &10.1  &   4000--9700   &  NOT/ALFOSC    \\
\hline 
\end{tabular}
\label{tab:speclog}
\end{center}
$^a$ Goodman High Throughput Spectrograph on SOAR.
$^b$ Robert Stobie Spectrograph on SALT. 
$^c$ Alhambra Faint Object Spectrograph and Camera (ALFOSC) on NOT.
\label{spectroscopic_log}
\end{table*}

\section{Comparison with Rapidly Evolving Transients from Photometric Surveys}
\label{sec:drout}

\begin{figure*}[htbp]
\centerline{\includegraphics[width=18cm]{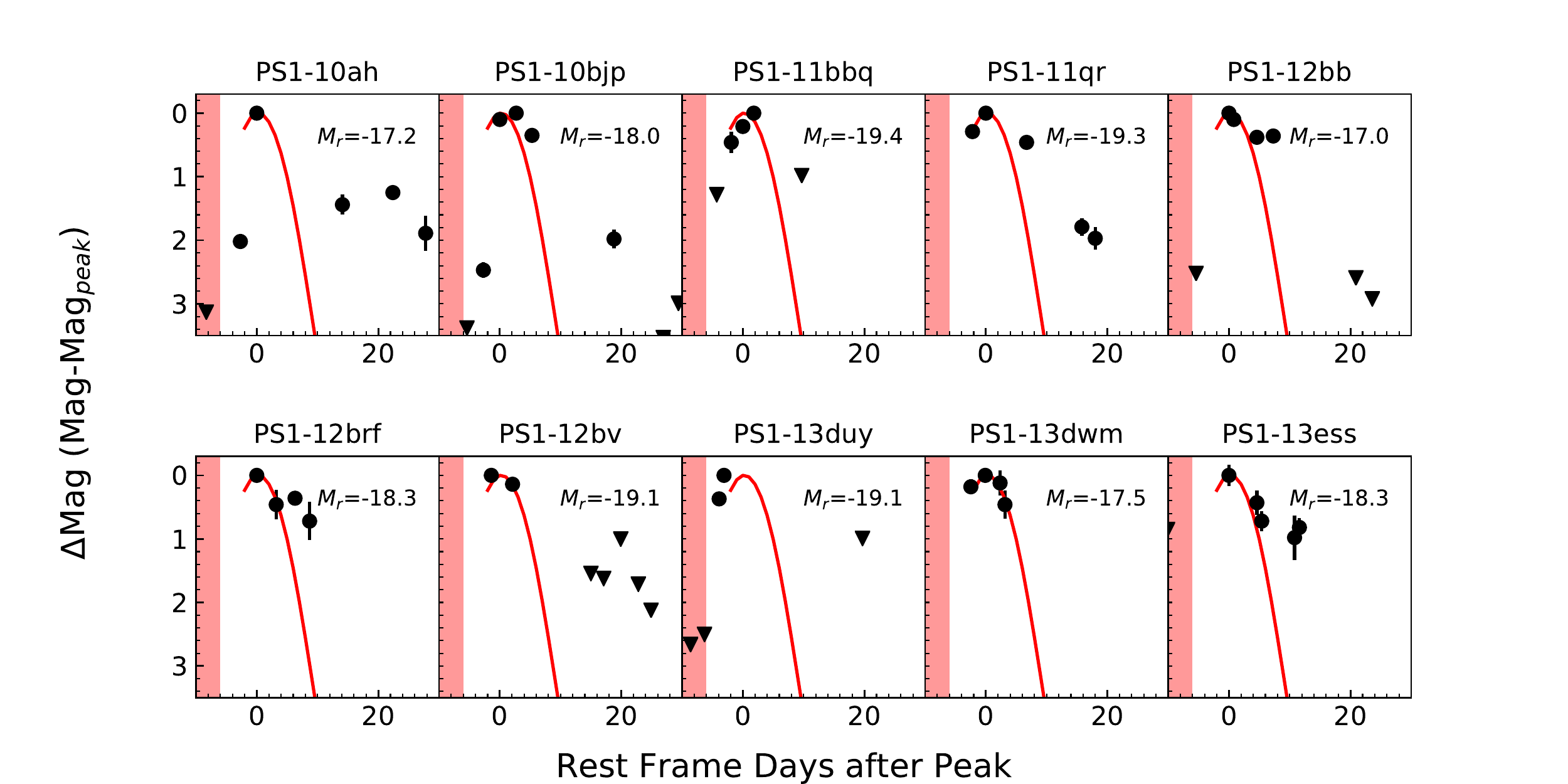}}
\caption{SDSS-r-band LC of SN 2019bkc (red line) compared to those of rapidly evolving transients from Pan-STARRS1 reported in \cite{Drout2014}. The estimated most likely explosion date of SN 2019bkc is within the vertical shaded region in red.}
\label{fig:drout14_compare}
\end{figure*}

\citet{Drout2014} presented a sample of 10 rapidly evolving and luminous transients with timescales above half maxima ($t_{1/2}$) of less than 12 days and $-20 < M_{\rm peak} < -16.5$ mag from a search within the Pan-STARRS1 Medium Deep Survey. Note that there were typically a few data points in the LCs, so there were no good constraints on peak time and brightness, and for each event, they adopted the observed brightest  epoch as the peak and measured $t_{1/2}$ by linearly interpolating the LC. The LCs of 2019bkc significantly distinguish from the sample of \citet{Drout2014} in the following aspects. First of all, SN 2019bkc declines faster than all objects with detections available at $\gtrsim10$ days for direct comparison. The measured decline in magnitudes between peak and 10 days post peak ($\Delta m_{10}$) for SN 2019bkc (listed in Table \ref{lc_paremeters}) are larger than all of the reported $\Delta m_{15}$ (decline between peak and 15 days post peak) for transients in \cite{Drout2014}. The $r$-band LC of SN~2019bkc is compared to all targets in the ``golden'' and ``silver'' sample of \cite{Drout2014}, as shown in Figure \ref{fig:drout14_compare}. PS1-13dwm has the most similar LC around the peak with 2019bkc, but it lacks data beyond 4 days after peak for further comparison. Furthermore, in all cases presented by \citet{Drout2014} with measured t$_{1/2}$ for rise and decline, the transients rise faster than they decline ($t_{1/2,\rm{rise}} < t_{1/2,\rm{decline}}$).  By adopting the same method to the SDSS $g$-band LC of SN 2019bkc, we obtain $t_{1/2,\rm{rise}}=5.3$ days, which is significantly larger than $t_{1/2,\rm{decline}}=2.3$ days, setting it apart from the transients of \citet{Drout2014}. Lastly, \citet{Drout2014}  found that their transients possess blue colors at peak ($g - r \lesssim -0.2$ mag). In contrast, SN 2019bkc has $g-r=0.04$ mag, which is substantially redder than most of the transients in \cite{Drout2014} and comparable to the reddest object PS1-12bb in their sample.         

\cite{Pursiainen2018} presented a sample of 37 rapidly evolving transients from a search within the Dark Energy Survey (DES) with a wide range of redshifts ($0.05<z<1.56$) and peak luminosities ($-22.25<M_g<-15.75$). To characterize the post-peak decay rates, \cite{Pursiainen2018} did exponential fits to measure the decay timescale $\tau$. The $g$-band decay timescales $\tau_g$ as shown in Figure 13 of \cite{Pursiainen2018} span from $\sim$2.5 to $\sim$15 days. The exponential decay timescale of SN 2019bkc in $g$-band is $1.94 \pm 0.07$ days, which is smaller than all objects in  \citet{Pursiainen2018}. SN 2019bkc is also redder (9000\,K at peak) than the majority of transients in \citet[][see their Figure 16]{Pursiainen2018}.

There are only a handful of (low-SNR) spectra for several transients presented in \cite{Drout2014} and \cite{Pursiainen2018}, and all of them show blue or red continua without clear detections of broad absorptions seen in SN 2019bkc.

\section{Explosion Environment}
\label{sec:environment}

Inspection of the Dark Energy Camera Legacy Survey \citep[DECaLS; ][]{Dey2019} $g$/$r$/$z$-band DR8 archival images (see Figure~\ref{fig:host}) reveals no detectable sources at the position of SN~2019bkc. It is projected onto, and likely associated with, the poor cluster MKW1 \citep[][shown in the left panel of Figure~\ref{fig:host}]{Morgan1975}, which has a velocity dispersion of $367^{+95}_{-54}$~km\,s$^{-1}$ \citep{Koranyi2002}. 
 The cD galaxy of MKW1 is NGC~3090 and other prominent cluster members include NGC~3086, 3092 (see the left panel of Figure~\ref{fig:host}),  and also NGC~3093, 3083, 3101 ($z=0.0204$, 0.0214, 0.0191, respectively; out of the view of the left panel of Figure~\ref{fig:host}). \citet{Koranyi2002} obtained a recessional velocity of $cz=6089\pm35$ km\,s$^{-1}$ for NGC~3090 and an average velocity of $\bar{cz}=6252\pm95$ km\,s$^{-1}$ for members within $0.5h^{-1}$ Mpc radius. 
 
\begin{figure}
\centerline{\includegraphics[width=15cm]{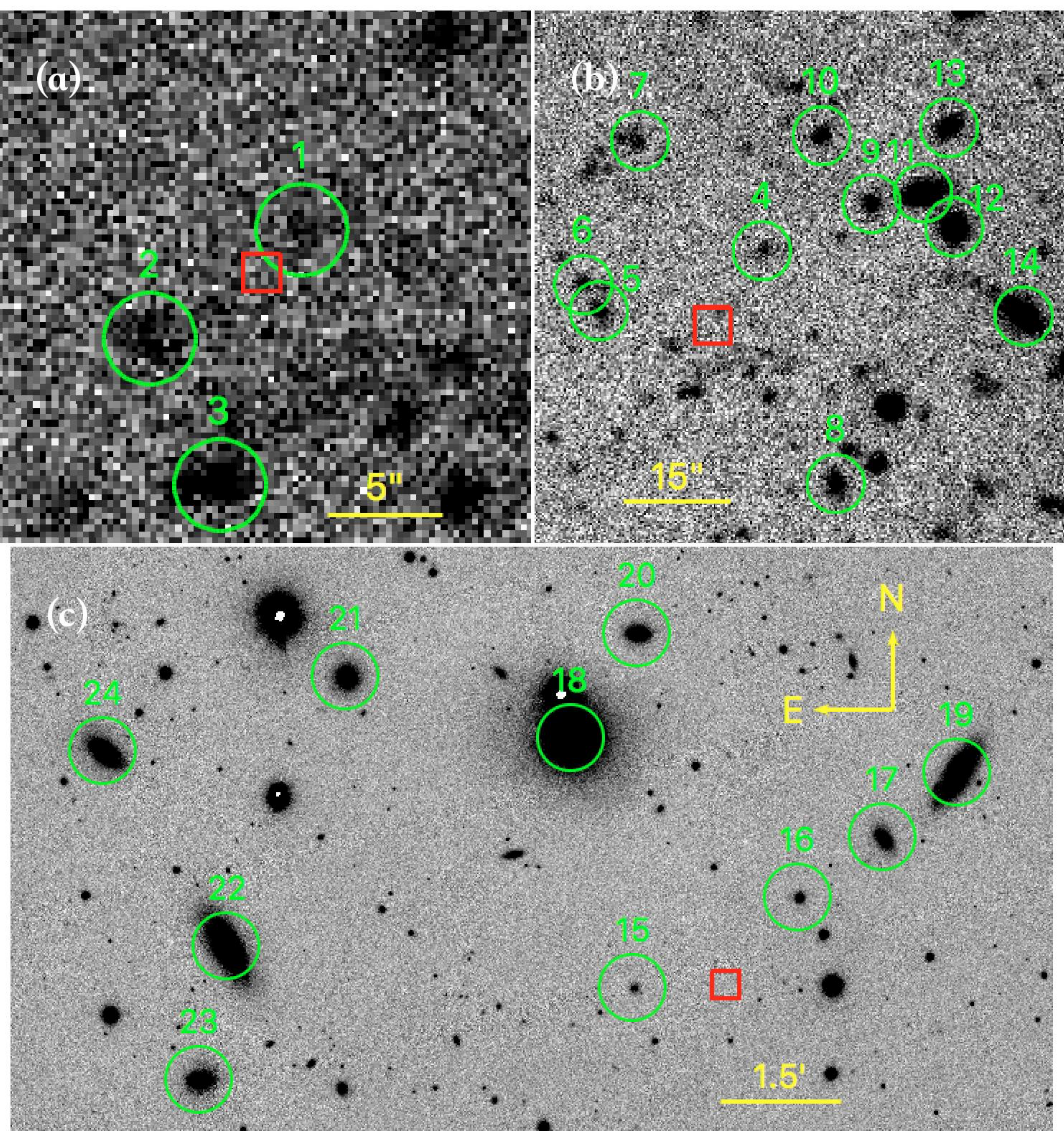}}
\caption{Explosion environment of SN 2019bkc, zoom-out views from panel (a)--(c). Images in panels (a) and (b) are stacked $g$-band images from NOAO Data Lab (see the text for information on the image) and image in panel (c) is $g$-band image from DECaLS DR8. The red square indicates the location of SN 2019bkc. The selected sources for chance alignment probability assessment are labeled with green circles, also listed in Table~\ref{tab:sources}. All of the labeled objects except object 1 are detected in DECaLS DR8, which are either spectroscopically confirmed galaxies (all labeled objects in panel (c) have spectra available in literature) or galaxy-like objects distinguished morphologically (all objects in panel (b) and object 2 and 3 in panel (a) have non-stellar-like morphologies from DECaLS DR8 catalog).  See the text for detailed discussion on object 1. }
\label{fig:sn19bkc_around_region}
\end{figure}

\begin{figure}
\centerline{\includegraphics[width=10cm]{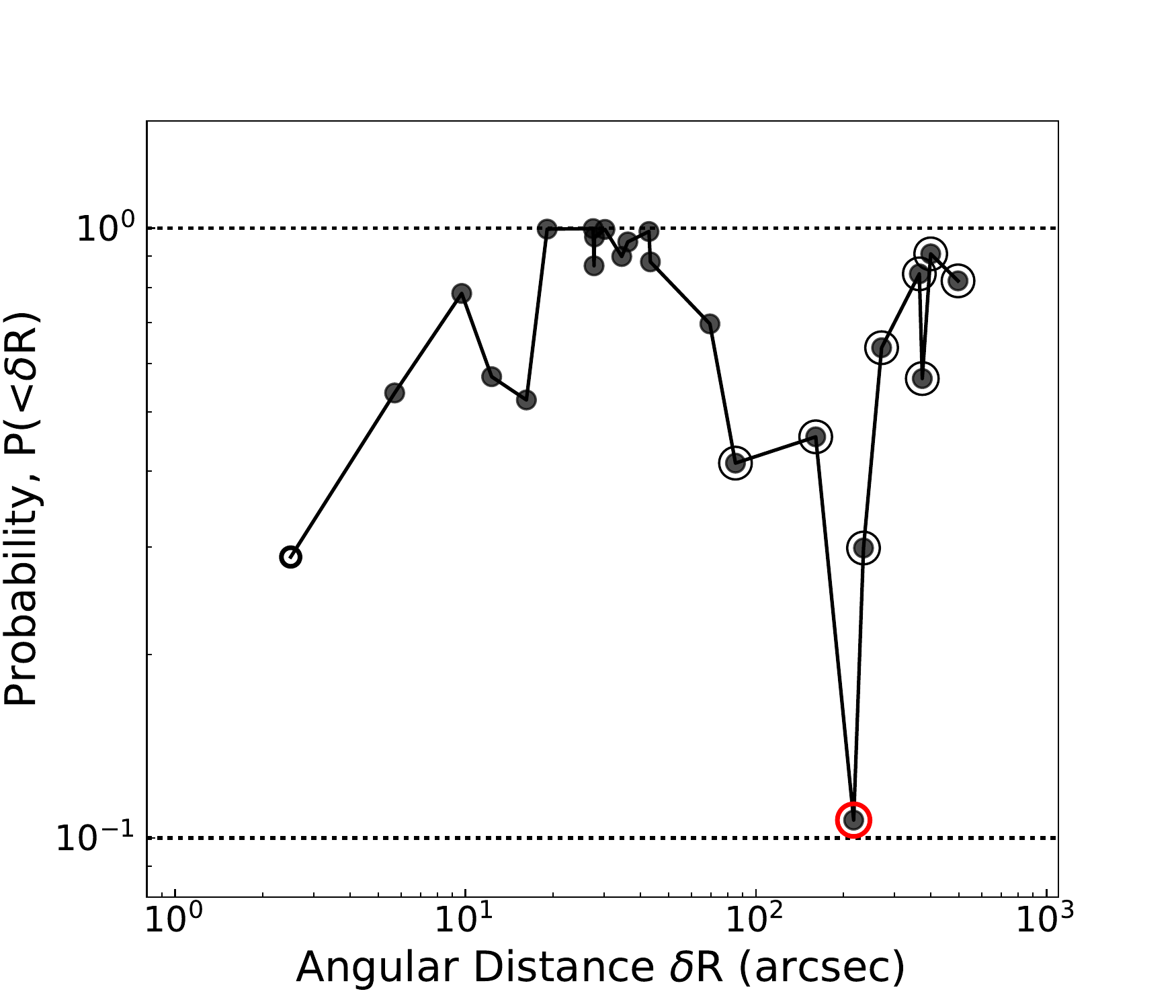}}
\caption{Probability of chance alignment as a function of angular distance to the explosion position with objects in Table~\ref{tab:sources}. The likely members of galaxy cluster MKW1 are marked with outer circles, and the one with lowest chance alignment probability, NGC 3090, is highlighted in red. The nearest object to SN 2019bkc cannot be distinguished from stellar object morphologically (see discussion in text) which is indicated by a circle.}
\label{fig:chance_probb}
\end{figure}

We assess the most likely host for SN 2019bkc by estimating the probability of chance alignment outlined in \cite{Bloom2002} and \cite{Berger2010}. The region around SN 2019bkc is shown in Figure~\ref{fig:sn19bkc_around_region} with the zoom-out view from panel (a) to panel (c). We calculate the probability of chance alignment with all labeled objects (listed in Table~\ref{tab:sources}) in Figure~\ref{fig:sn19bkc_around_region}  based on their $r$-band magnitudes. All objects in Table ~\ref{tab:sources} except object 1 are detected in DECaLS DR8, which are either spectroscopically confirmed galaxies or galaxy-like objects distinguished by morphology. The result is shown in Figure~\ref{fig:chance_probb}. The cD galaxy of MKW1, NGC 3090, has the lowest probability of chance alignment with SN 2019bkc ($\sim$10\%) and another three galaxies belonging to MKW1 have probability of chance alignment $<$ 50\%, which strengthen our argument for the likely association of SN 2019bkc with cluster MKW1. 
 
The projected distance of  SN~2019bkc from NGC~3090  is 88.6\,kpc (see Figure~\ref{fig:host}). It is about 34\,kpc from the closest known cluster-member galaxy 2dFGRS N216Z084 ($z=0.019$). SN~2019bkc is therefore likely located in the outer  halo or is an intracluster source, where no recent star formation activity is expected to have occurred. 
 
 We also retrieved deep images in $g$- and $i$-band from NOAO Data Lab\footnote{\url{https://datalab.noao.edu/}} covering the field of SN~2019bkc. The images have been obtained with DECam on CTIO-4m telescope under program 2017A-0060.  The total exposure times are 2280s and 1015s for the $g$-band and $i$-band stacked images, respectively. There are no visible sources at the position of SN 2019bkc with 3$\sigma$ detection upper limit of $m_g\approx 26.2$~mag and $m_i\approx 24.8$ mag (see top-left panel of Figure~\ref{fig:sn19bkc_around_region}). The closest source to SN~2019bkc, object 1 in Table~\ref{tab:sources}, is 2$\farcs$5 away from SN 2019bkc on the projected sky, and it is only detected on $g$-band image for which we get $m_g = 25.7\pm0.3$ from aperture photometry with an aperture radius of 1$\farcs$6. The source is not spatially resolved with seeing $\sim 1\farcs4$, and its low S/N prohibits us from studying the morphology of source to securely distinguish it between star and galaxy. It is noted that the color of object 1 ($g-i<0.9$) is bluer than those of all MKW1 galaxies.  
 
The $3\sigma$ $g$-band detection limit for the stacked DECam image of $m_g\approx26.2$~mag, corresponds to $M_{g} \gtrsim -8.5$. At the distance of MKW1, 1\farcs5 = 820~pc, which means that a typical low-luminosity dwarf galaxy will have a size comparable to the seeing scale. From the Local Group dwarf galaxy compilation of \citet{McConnachie2012}, we find that nearly every gas-rich (i.e., star-forming) Local Group dwarf would be brighter than this limit only with Leo~T ($M_V=-8$ mag) as an exception. 
This absolute magnitude limit cannot rule out all but the most luminous GCs  \cite[see e.g.,][]{Harris1996}.

There is the possibility that SN~2019bkc resides in an isolated star forming region. \ion{H}{2} regions in galactic outskirts are rare, but have been found associated with tidal \ion{H}{1} features in the outskirts of gas-rich galaxies \citep{RyanWeber2004}. Although MKW1 does not appear to be a gas-rich environment, we can constrain the possibility of SN~2019bkc being in an isolated star-forming region. None of our spectra show discernible H$\alpha$ emission at the location of the SN, with an upper limit on the line flux of $\sim\!10^{-17}$~erg~s$^{-1}$~cm$^{-2}$ from the NOT/ALFOSC spectrum taken at $+10.1$ days. At the assumed distance of the SN, this translates into a maximum H$\alpha$ luminosity of $\sim\!10^{37}$~erg~s$^{-1}$, or a star formation rate of less than $10^{-5} M_\odot\ {\rm yr^{-1}}$. This is comparable to the luminosity of the Orion Nebula, whose ionizing flux is dominated by a single O star \citep{Pellegrini2012}. 
Although almost all ccSNe are associated with star forming regions, there are some notable exceptions such as SN~2010jp \citep{Smith2012}. This peculiar SN~II   was 33\,kpc from its host center, and there was no identifiable source in a pre-explosion image to a limiting absolute magnitude of $\sim -12$.

A runaway star is a possible origin for a remote, massive star. However, given the $\sim 30$\,Myr lifetime of a $\sim 10\,M_\odot$ star, it would require a velocity $\gtrsim 1000$\,km/s to travel the projected distance of $30$\,kpc to the closest member galaxy. Such a velocity is comparable to that of the fastest known hypervelocity star \citep[e.g.,][]{Brown2015}, which makes it an unlikely scenario. 

We conclude that the lack of any obvious star-forming host galaxy and the non-detection of H$\alpha$ in the late-time spectrum make a massive star progenitor relatively unfavorable.

\begin{table*}[htbp]
\caption{Catalog of Sources in the Field of SN 2019bkc}
\begin{center}
\begin{tabular}{cccccccc}
\hline
\hline
 ID      &   Name           &      $z$       &   Angular Distance &   $m_g$ & $m_r$     & $m_i$   & Chance Alignment Probability   \\
           &                       &             &   (arcsec) &   ( mag) &  (mag)     &   (mag)  &    \\
 \hline
 1        & Uncatalogued &   n/a       &   2.5                      &   25.7     &  $\cdots$ &$>$24.8&    0.289  \\ 
 2        & Uncatalogued &   n/a       &   5.7                      &   24.7     &  23.7       &$\cdots$&    0.537  \\        
 3        & Uncatalogued &   n/a       &   9.7                      &    24.3    &   24.1       & $\cdots$&    0.782 \\
 4        & Uncatalogued &   n/a       &   12.3                     &  24.4      &   22.7       & $\cdots$&    0.571      \\
 5        & Uncatalogued &   n/a       &   16.2                     &   23.5     &   21.8       &   $\cdots$      &    0.523     \\   
 6        & Uncatalogued &   n/a       &   19.1                     &   24.1     &   24.1       & $\cdots$&    0.997     \\
7        & Uncatalogued &   n/a       &    27.5                     & 24.2       &    24.0     &  $\cdots$& 0.999  \\
8        & Uncatalogued &   n/a       &    27.7                      & 22.9      &   21.3      &  $\cdots$    &  0.868 \\    %
9        & Uncatalogued &   n/a       &     27.8                     & 23.1      & 22.4         &   $\cdots$         &   0.968\\
10        & Uncatalogued &   n/a      &    30.2                     & 23.3       & 22.8        &   $\cdots$         &   0.996\\
11& SDSS J100020.59-030054.6 &   n/a      &    34.5                    &  21.9       & 21.2         &   21.6         &   0.899\\   
12& SDSS J100020.27-030059.3&   n/a      &    36.5                    &  22.0        & 21.5        &    21.0         &  0.950 \\
13        & Uncatalogued &   n/a      &    42.8                    &  22.6       &  21.6         &    $\cdots$         &  0.988 \\    
14& SDSS J100019.65-030111.1 &   n/a      &    43.3                    & 21.4       & 20.6          & 20.0            &   0.881\\ 
15 &2dFGRS N216Z080 & 0.1880   &  69.4 &  18.6 & 17.6 &  $\cdots$  & 0.697\\
16 &2dFGRS N216Z084 & 0.0190   &  85.0 & 17.8  & 17.0 & 16.6&0.412 \\
17 & 2dFGRS N216Z089 & 0.0217   & 160.6 & 16.3  & 15.5 & 15.1 & 0.455\\
18         &NGC 3090        &  0.0203  &  217.0                   &  13.5      & 12.5       & 12.1      &    0.107      \\  
19         &NGC 3086        &  0.0225  &  234.6                   &  14.6       & 13.8        & 13.3     &   0.299\\
20      & 6dFGSv 05125  &0.0223    &   270.8 & 15.6 & 14.8 & 14.3 & 0.637\\
21& 2dFGRS N216Z072 & 0.0202    & 365.2  & 15.7  & 14.8 & 14.2 & 0.842\\
22        &NGC 3092        &  0.0197    &  373.9                   &   14.5      &  13.7       & 13.2    &    0.567 \\
23 & 2dFGRS N217Z145 & 0.0190   & 399.7                    & 15.7  & 14.9  & 14.5 & 0.908\\
24        & NGC 3093        & 0.0204    & 496.3  & 14.8 & 13.9 & 13.4 & 0.820\\
\hline 
\end{tabular}
\label{tab:sources}
\end{center}
\label{spectroscopic_log}
\end{table*}


\section{$^{56}$Ni Contribution to the Bolometric LC of SN 2019bkc}
\label{sec:ni56}

\begin{figure}
\centerline{\includegraphics[width=18cm]{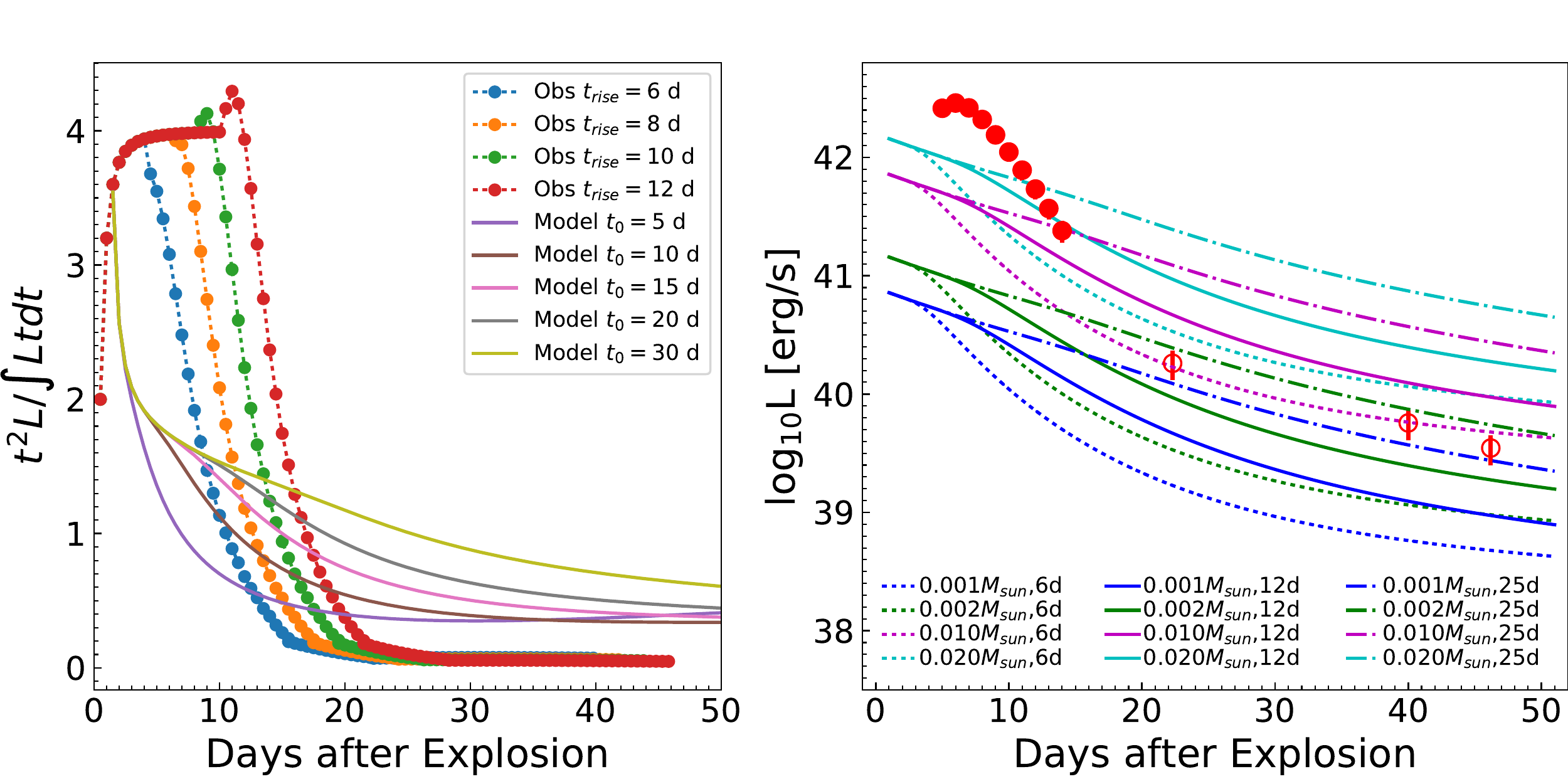}}
\caption{Left panel: models of radioactive deposition from $^{56}$Ni decay with different $\gamma$-ray escape time-scale $t_0$ shown as solid lines compared with observed bolometric light for SN 2019bkc adopting a range of explosion times shown as circles with connecting dots. Right panel: constraints on $^{56}$Ni mass assuming the late-time tail of the bolometric LC, shown as red circles, is powered by $^{56}$Ni decay. The $^{56}$Ni mass and  $\gamma$-ray escape time-scale are given for each model in legend. Here $t_{\rm rise}=6$ days is used. }
\label{fig:ni56_test}
\end{figure}

If the transient is powered by $^{56}$Ni decay, like in SNe Ia, the radioactive-decay energy is principally released
as $\gamma$-rays and positrons. The $\gamma$-ray deposition fraction is time-dependent and can be approximated as $1-\exp{(-t_0/t)^2}$, where $t_0$ is the $\gamma$-ray escape time-scale. As shown in \cite{Katz2013} and \cite{Wygoda2019}, at late-time $t\gg t_{\rm peak}$ when there is negligible trapped optical radiation in the ejecta, there are two relations between the deposition energy and the bolometric luminosity, $Q(t)=L(t)$ and the integral $\int_0^tQ(t')t'dt' = \int_0^tL_{bol}(t')t'dt'$. We first linearly interpolated the bolometric LC shown in Figure \ref{fig:Lbol_compare} for the late times, then extrapolated the bolometric LC to the explosion time in a quadratic way. The time from explosion to $B$-band peak $t_{\rm rise}$ is uncertain and we explored a wide range of 6--12 days. In the left panel of Figure~\ref{fig:ni56_test}, ${t^2L}/{\int Ltdt}$ values derived from our observed data (shown as circles with connecting dots) fall rapidly and flatten to $\lesssim$0.2 at phases $\gtrsim$ 20 days after explosion. The $^{56}$Ni decay models (shown as solid lines) fail to fit the data at late time ($\gtrsim$ 20 days) even though a wide range of the $\gamma$-ray escape time-scale $t_0$ from 5 days to 30 days is allowed. The significant mismatch between $^{56}$Ni decay model with observations suggests that SN 2019bkc cannot be dominantly powered by radioactive decay of $^{56}$Ni.

However, the late-time tail of SDSS-$i$-band LC of SN 2019bkc in Figure~\ref{fig:lcs_compare} may be powered by  $^{56}$Ni radioactive decay. We explore to what extent $^{56}$Ni contributes to the radiation energy of SN 2019bkc.  Radioactive decay energy deposition models with a range of values for $^{56}$Ni mass and  $\gamma$-ray escape time-scale are shown in the right panel of Figure~\ref{fig:ni56_test} where $t_{\rm rise}=6$ days is adopted. We restrict $t_0\gtrsim6$ days because it is not physical to have a $\gamma$-ray escape time-scale shorter than $t_{\rm rise}$. $t_0$ and $M_{^{56}Ni}$ are degenerate in producing a late-time tail that is consistent with observations. With $t_0\gtrsim6$ days, we get an upper limit for $^{56}$Ni mass of $M_{^{56}\rm Ni} \lesssim 0.01 M_{\astrosun}$, which can only account for $\sim$10 \% of the peak luminosity of SN 2019bkc. The estimated $M_{^{56}\rm Ni}$ by fitting the tail is between 0.001 and $0.01\,M_{\astrosun}$.

\section{Explosion models and progenitor scenarios to explain SN 2019bkc}
\label{sec:comparemodels}

SN 2019bkc provides new and significant constraints on existing models with the key properties summarized in \S~\ref{sec:summary}. The luminosity constraint in the following used to distinguish explosion models is subject to the adopted redshift $z=0.0209$ for SN~2019bkc. Below we examine various explosion models and compare them with SN~2019bkc. 

One proposed class of progenitor systems includes at least one low-mass compact star (for example, a WD), which is consistent with the likely old stellar population environment of SN 2019bkc. \citet{Shen2010} studied the  SNe~.Ia produced by the He detonation of a WD due to materials accreted from a He-star companion. They adopted WD masses $M_{\rm WD}$ ranging from 0.6 to 1.2~$M_{\odot}$ and envelope masses $M_{\rm env}$ from 0.02 to 0.3~$M_{\odot}$. Their predicted LCs are shown as black solid lines in Figure~\ref{fig:Lbol_compare}.  While some of the models can reach the peak luminosity of SN~2019bkc, they all decline too slowly compared to SN~2019bkc.  The models that decline more rapidly have peak luminosities that are too dim, which would require a closer distance of SN~2019bkc than the adopted value to match with observations. Furthermore, the rise times from these models are also too fast compared to SN~2019bkc. Next we examine the model by \cite{Darbha2010}, which studied the properties of LC and spectra from nickel-rich outflows produced by AIC. They found a peak luminosity of $2\times10^{41}{\,\rm erg\,s^{-1}}$ ($5\times10^{40}{\,\rm erg\,s^{-1}}$) only $\lesssim$1 day after the collapse for an ejecta of $10^{-2}\, M_{\odot}$ ($3\times10^{-3}\, M_{\odot}$) and the decay time is 4 (2) days. The bolometric LCs are shown as green lines in Figure~\ref{fig:Lbol_compare} with the representative electron fraction (0.45 -- 0.55). The bolometric LCs of transients powered by the radioactive heating of low-mass ejecta are faster and more than one order of magnitude dimmer compared to SN 2019bkc, though the latter could be alleviated if 2019bkc were at a substantially closer distance. Furthermore, the near-peak spectra from their calculation are dominated by significantly Doppler-broadened Ni features with no distinct absorption spectral lines present, and they are markedly different from observed spectra of SN 2019bkc. Therefore, the specific AIC model in \cite{Darbha2010} fails to reproduce the LCs and spectra of SN~2019bkc. 

Another class of progenitors involves a massive star. \citet{Tauris2013a} studied ultra-stripped SNe Ic from close-binary evolution and reproduced the bolometric LC of SN~2005ek with ejecta mass $M_{\rm ej}=0.2\, M_{\odot}$, explosion energy $E_{\rm ej}=5\times10^{50}$ erg and synthesized $^{56}$Ni mass $M_{\rm Ni}=0.05\, M_{\odot}$. As shown by the yellow solid line in Figure~\ref{fig:Lbol_compare}, its predicted peak luminosity is similar to SN~2019bkc, but the post-peak decline rate is much slower than SN~2019bkc, and the late-time LC is significantly more luminous compared to that of SN 2019bkc. For ultra-stripped SNe Ic, the short time scale around peak is due to low ejecta mass, which results in short diffusion time and enhanced $\gamma$-ray escape. \citet{Tauris2015} further generalized the work to a larger variety of ultra-stripped SNe and studied the light-curve properties for explosions including more extreme low-mass ejecta (see their Table 2), which can achieve more rapid evolution. In essence, the LCs of ultra-stripped SNe Ic models are mainly powered by $^{56}$Ni decay, which is inconsistent with 2019bkc. It is also worthy to note that the decay time is generally longer than rise time for ultra-stripped SNe in \citet{Tauris2015}, which does not agree with SN 2019bkc. 
\citet{Moriya2016} studied rapidly evolving dim transients from stripped-envelope ECSNe. In Figure~\ref{fig:Lbol_compare}, we display the full collection of their models (type~Ibc ECSN bolometric LCs with explosion energy $1.5\times10^{50}$ erg, $^{56}$Ni mass $2.5\times10^{-3}\, M_{\odot}$, typical ejecta mass 0.3\,M$_{\odot}$ from metallicity $Z=0.02$ progenitors). In this scenario, the transient is powered by  $^{56}$Ni decay, and the predicted yield is broadly consistent with the late-time tail of SN~2019bkc but too dim compared to the peak at the adopted distance.

\citet{Dessart2011} presented fast-evolving ccSNe models from Wolf--Rayet progenitors when no $^{56}$Ni is produced
(models Bmi25mf6p49z1, Smi60mf7p08z1, Bmi25mf7p3z0p2, and Smi25mf18p3z0p05 in their Tables~1 and 2). For these models, the final pre-explosion masses range from 6.49 to 18.3  $M_{\odot}$, and result from the evolution of  25 $M_{\odot}$ or 60 $M_{\odot}$ zero-age-main-sequence stars. As shown in in magenta in Figure~\ref{fig:Lbol_compare}, the LCs start with shock breakout and then enter shock cooling phases. Some of the shock cooling LC can achieve decline rate comparable to that of SN 2019bkc, while the peak luminosities of shock cooling phase are too low, which would require significantly closer distances than the adopted value to match with the observed LCs. The peak luminosity and evolution of shock cooling LC are mainly controlled by the ejecta kinetic energy and the stellar radius. For Wolf--Rayet progenitors, the final stellar radii that they adopted are small (0.45--3.64 $R_{\odot}$ for their $^{56}$Ni deficient models) where energy deposition by shock is insufficient. Furthermore, such models cannot reproduce the observed tail due to no $^{56}$Ni production.

\citet{Kleiser2018b} studied explosions of 5 $M\odot$ hydrogen-free massive stars with circumstellar shells where little or no radioactive nickel is ejected, and the source of radiation is almost all due to shock cooling.  They explored a large parameter space of progenitor/explosion properties including shell mass $M_{\rm shell}$, density profile (given in gaussian form which is specified by the radius of peak density $r_{\rm mid}$, mass-loss velocity fixed to 100 km s$^{-1}$, and mass-loss duration $\tau$) of the shell and explosion energy $E_{\rm exp}$. One of the LC shown in Figure~\ref{fig:Lbol_compare} ($M_{\rm shell}$=1.0 $M_{\odot}$, $r_{\rm mid}=1\times10^{12}$cm, $\tau$=10 d, $E_{\rm exp}=6\times10^{51}$erg) has similar peak luminosity and decline rate with SN 2019bkc, while we note that the extreme explosion energy and arbitrary wind-like mass shell profile has been used in the calculation. \citet{Kleiser2018a} produced stellar models with varying zero-age-main-sequence mass $M_{\rm ZAMS}$ between 12 and 18 $M_{\odot}$ such that their bare helium cores lie in the 2--4 $M_{\odot}$ range once the hydrogen envelope is removed. Strong He and C shell burning outside of a core will develop radii as large as 100 $R_{\odot}$ which can naturally result in bright and rapid declining shock-cooling LCs. Two LCs from such a helium giant star explosion are shown in Figure~\ref{fig:Lbol_compare} with explosion energy of $1\times10^{51}$ erg for both and star masses of 2.01 and 2.26 $M_{\odot}$. The peak luminosities and post-peak decline rates are roughly consistent with SN 2019bkc. 
It is shown that a small amount of $^{56}$Ni can produce a tail as seen in SN 2019bkc while most of the peak luminosity still comes from shock cooling by \cite{Kleiser2018a, Kleiser2018b}, which is also demonstrated in Appendix~\ref{sec:ni56}. Also, evolution with slower rise than decay around peak LC are commonly seen in shock cooling LCs which is consistent with what we see in SN 2019bkc. Energetic explosion from expanded helium stars with small amount of $^{56}$Ni yield can reproduce the LC of SN 2019bkc. The thick helium envelop is not disfavored by the tentative helium detection as discussed in Appendix~\ref{sec:spec_features}. However, the likely old stellar environment of SN 2019bkc makes it challenging for a massive star progenitor to explain SN 2019bkc.

\end{document}